\documentclass[twocolumn,fleqn,usenatbib]{mnras}

\usepackage{graphicx}
\usepackage{color}
\usepackage[nointegrals]{wasysym}
\usepackage{mathtools}
\usepackage{subcaption}
\usepackage{orcidlink}
\usepackage{booktabs}

\usepackage{newtxtext,newtxmath}
\usepackage[T1]{fontenc}
\usepackage{bm}
\usepackage{amsmath}
\usepackage{graphicx}	
\usepackage{amsmath}	

\newcommand{\ms}[1]{{\color{black}#1}}

\title[The helium triplet in outflows]{Using the helium triplet as a tracer of the physics of giant planet outflows}

\author[M. Schulik and J. E. Owen]{
Matth\"aus Schulik$^{\orcidlink{0000-0001-6460-0759}1,2}$\thanks{E-mail: mschulik@ic.ac.uk (Imperial)}
and James E. Owen,$^{\orcidlink{0000-0002-4856-7837}1,2}$
\\
$^{1}$Imperial Astrophysics, Department of Physics, Imperial College London, Prince Consort Rd, London SW7 2AZ, UK\\
$^{2}$Department of Earth, Planetary, and Space Sciences, University of California, Los Angeles, CA 90095, USA}

\date{Accepted XXX. Received YYY; in original form ZZZ}

\pubyear{2015}

\begin{document}
\label{firstpage}
\pagerange{\pageref{firstpage}--\pageref{lastpage}}
\maketitle

\begin{abstract}
Hydrodynamic outflows, such as those observed escaping close-in gas giant planets, are not isothermal in structure. Their highly ionized nature allows them to cool adiabatically at distances beyond several planetary radii. The contrast between the hottest gas temperatures at around 10,000K and the coldest at around 1,000K triggers an excess population of the observable helium triplet. This excess is caused by the suppression of collisional de-excitation from the triplet state at cool temperatures.  Using radiation-hydrodynamic simulations, we show that this helium triplet excess may explain the excess broadening seen in HD 189733b's observed transmission spectrum, demonstrating adiabatic cooling of its outflow, confirming its hydrodynamic nature on scales of several planetary radii. However, further observations are required to confirm this conclusion. Furthermore, we explore a range of electron transitions for neutral helium which were not considered in the previous literature. We find that the He$2^1$S state is unavailable as a potential reservoir for He$2^3$S electrons. Additionally, the de-excitation to the ground state must be considered for stellar spectra later than K2 in predicting the correct helium triplet population. Importantly, since triplet helium inherits momentum from ionized helium as it is generated by recombination, it is significantly less prone to fractionation than ground-state neutral helium. However at separations of $\gtrsim 0.05$~au, ionization at the flow base and drag on helium weaken, leading to significant fractionation of the then mostly neutral helium. This in turn, can cause a suppression of the Helium transit depth, even though the helium line width remains large.
\end{abstract}

\begin{keywords}
Planets and satellites: atmospheres -- Planets and satellites: detection 
\end{keywords}



\section{Introduction}
\label{sec:introduction}

Atmospheric escape is believed to play an important role in sculpting the atmospheres of planetary bodies across a wide range of parameters. Atmospheric escape has been measured in the terrestrial planets of the Solar System in various species such as atomic hydrogen and oxygen, constraining their histories \citep{lammer2021}. In the case of close-in exoplanets, the extreme irradiation they experience heats up their upper atmospheres causing powerful atmospheric escape in the form of hydrodynamic winds \citep{yelle2004,garciamunoz2007,MurrayClay2009, Owen2012}.  Transmission spectroscopy of exoplanets has provided evidence for atmospheric escape from close-in planets in the Lyman-$\alpha$ line \citep{vidal-madjar2004, lecavelier2012}. Recently, detections in the helium 10830~\AA~ triplet line \citep{Oklopcic2018, Spake2018, dosSantos2023}, and UV metal lines \citep[e.g.][]{Sing2019} have extended our ability to observationally study atmospheric escape from exoplanets,  with the latter explicitly demonstrating the hydrodynamic nature of the outflow.
In addition, measurements of escaping exoplanetary atmospheres can potentially tell us about their atmospheric compositions and planetary magnetic fields \citep{kislyakova2014, Schreyer2024}. 

Understanding the details of atmospheric escape from exoplanets is critical to disentangling the impacts of formation, thermal evolution and mass-loss. This need is particularly pertinent because the observed exoplanet population is typically billions of years old \citep{mcdonald2019,Petigura2022}, where evolutionary processes have already sculpted its properties. A mass-loss driven evolutionary history has had great success in explaining many demographic features in the exoplanet population \citep[e.g.][]{Owen2019review}, both in the context of photoevaporation \citep[e.g.][]{owen2017,rogers2021} and core-powered mass-loss \citep[e.g.][]{Gupta2019}. However, different assumptions about the mass-loss rates lead to different conclusions about the exoplanet population's initial properties \citep[e.g.][]{rogers2021} and their associated formation pathways \citep[e.g.][]{OA19}.  

Recently, alternative mechanisms to create the demographics features, such as the radius gap, have been proposed, such as formation-based scenarios \citep{Venturini2020,lee2022, Burn2024, Venturini2024} or atmospheric removal via impacts \citep{zhong2024}. Therefore, while observations demonstrate that atmospheric escape from close-in exoplanets occurs, more quantitative tests of these escape models used in evolutionary calculations are required. Given the helium 10830~\AA~ triplet line's success as the now leading atmospheric escape tracer (in terms of pure detection number, \citealt{dosSantosreview}), a detailed understanding of its formation and information content is required. 




In this work, we deliver an update on some important helium population rates, and we propose a test of photoevaporation theory to ascertain or refute its validity. Given giant planets' large sizes, which result in high signal-to-noise observations, the most sensible planetary mass range to begin these quantitative tests is the escaping atmospheres of giant planets.

Giant planets possess high surface gravities, out of which atmospheric gas has considerable difficulty escaping. On one hand this might lead to significant fractionation of heavier elements, such as helium, as has been inferred in \cite{lampon2021} and \cite{xing2023}.
On the other hand, those flows spending considerable time in the gravity well of their planets \citep{Caldiroli2021}, giving them time to become strongly ionized \citep{MurrayClay2009}, enhancing the helium signal \citep{Oklopcic2019}. As giant planet upper atmospheres are necessarily mostly atomic hydrogen, their ionized versions, i.e. protons, have neither strong heating nor cooling properties. Thus, they are thermostated to temperatures around $10^4$K by Lyman-$\alpha$ cooling from the small population of neutral hydrogen generated by recombination \citep{MurrayClay2009,owenalvarez2016}.
Hence, after being heated to $\sim 10^4$K, and the densities have dropped sufficiently that Lyman-$\alpha$ cooling becomes inefficient, the escaping ionized giant planet atmospheres have only one simple cooling pathway - that of adiabatic expansion \citep{MurrayClay2009}.
This mechanism is well understood and can deliver temperature drops down to $\leq 10^3$K. Since the helium triplet is typically an optically thin tracer whose signal can be dominated at radii of a few planetary radii \citep{linssen2023, ballabio2025}, this temperature drop can be probed observationally. 
An important consequence of this temperature drop is the production of a local maximum in the observable helium triplet, as already pointed out by \citet{yan2022} and \citet{biassoni2024}. {Since the triplet maximum around giant planets is situated at only a few planetary radii, those signals can be interpreted the easiest and clearest without the complicating interference of the stellar wind, which is the case for some lower-mass planets \citep[e.g.][]{Zhang2023}. Due to these interactions, the helium signal could be heavily modified \citep[e.g.][]{wang2021, wang2021b} or erased as the planetary outflow is crushed by the stellar wind. Thus, the absence of a population maximum would not invalidate photoevaporation but would merely add to the already existing non-detections potentially arising from stellar winds or planetary magnetic fields \citep[e.g. ][]{Schreyer2023,Alam2024}.}

\ms{The unambiguous detection of an adiabatic cooling signature associated with such a triplet population maximum would be a powerful indicator that photoevaporation theory is correct.} Therefore, we investigate what properties this population maximum might have on our observational signatures of the helium line, in conjunction with the proposed detections of fractionated helium. In particular, we focus on its impact as an additional broadening mechanism for the triplet line beyond thermal broadening, and we further investigate the kinematic signatures of all\ms{, potentially fractionating} helium species involved in more detail.

\section{Methods}

On the largest scales, atmospheric outflows interact with the circumstellar environment, creating complex 3D structures \citep[e.g.][]{bourrier2013, kislyakova2014, khodachenko2019}.  In the case of helium triplet observations, 3D simulations have been used to interpret the tails \citep[e.g.][]{Macleod2022,Nail2024} that have been observed in a few cases \citep[e.g.][]{Spake2021,Zhang2023}. However, these 3D simulations are extremely computationally intensive. Given the majority of the helium signal originates inside the planet's hill sphere, where the outflow remains approximately spherically symmetric \citep[e.g.][]{MurrayClay2009,Tripathi2015} we use the standard approach of solving 1-D spherically symmetric hydrodynamic equations, coupled to a calculation of the neutral helium level population  \citep[e.g.][]{Oklopcic2018,biassoni2024,allan2024}. 
Since our focus in this work is on giant planets, 
and studies of the detailed formation of the helium triplet signal show that the signal is formed close at a few planetary radii
\citep{linssen2023}, a 1-D approach is the natural starting point. Thus, we use {\sc aiolos} to solve the 1-D radiation hydrodynamics equations \citep{schulik2023}, and modify this code to include the physics necessary to simulate the helium triplet.

\subsection{Processes included}
\citet{Oklopcic2018} presented a simplified atomic level population calculation for determining the helium triplet $2^3$S level population, highlighting the key physics. \cite{yan2022} and \cite{biassoni2024} pointed out the importance of the non-isothermal treatment of all coefficients in the helium level population calculation. These \emph{temperature dependent} rates become particularly important for the escaping outflows from giant planets. This is because their high masses result in large temperature variations provided by adiabatic cooling \citep{salz2016}, as the outflow expands. Unlike atmospheric outflows from lower mass planets, e.g. sub-Neptunes, giant planets typically have escape temperatures above $10^4$~K. Therefore, their strong gravities result in small-scale heights and, thus, strong heating near the planet. Once the atmosphere reaches $\sim 10^4$~K, it expands rapidly, cooling through PdV work well below this temperature.
\ms{Reheating after adiabatic cooling is prevented as the ionization -recombination equilibrium is maintained at large planetocentric distances, starving the gas of neutral hydrogen, which can absorb ionizing photons, heating the gas \citep{MurrayClay2009}.}
Thus, hot Jupiters around main-sequence stars, the easiest to study observationally, are also those most likely to demonstrate significant PdV cooling, making them ideal targets. Recently, \citet{allan2024} studied the evolution of the triplet population in planetary outflows driven by K-type stars of different ages. Notably, they expanded the helium level calculation to include the $2^1$S state and additional processes, such as charge exchange, ionisation and recombination into all states, finding significant build-up of $2^1$S electrons. 

In our \texttt{aiolos} simulations we include the relevant processes listed in \citep{allan2024}, i.e. for the levels of neutral helium: $1^1$S, $2^3$S, $2^1$S, $2^1$P, and $\rm He^{+}$, $\rm He^{++}$, such as recombination, Penning ionization, radiative decay, electron-collision induced transitions and charge exchange. We include multiband ionization due to various stellar spectra. Our high-energy radiative scheme uses 22 bands from 24nm to 260 nm, with the band limits adjusted to discontinuities in the opacity functions of all species.  Several previous works \citep[e.g.][]{p-winds} assumed isothermal outflows at $10^4$~K, which are often used to interpret observations \citep[e.g.][]{Allart2023,Alam2024}, and thus neglected the temperature dependence of the collision rates; given they are exponentially sensitive to temperature, they play an important role in setting the helium triplet population \citep{yan2022, biassoni2024, ballabio2025}. In particular,  \citet{biassoni2024} pointed out the importance of using the rate $r^{\uparrow}_{23}$ ($q_{31a}$ in the notation of \citealt{Oklopcic2018}) as temperature-dependent quantity.
 As the helium triplet state is typically depopulated by electron collisions, especially for those outflows that are observable \citep[e.g.][]{Oklopcic2019}, lower temperatures slow the rate of depopulation. This slower depopulation rate leads to the decrease of this main loss channel for He$2^3$S, and hence, a strong radial increase in the fraction of helium atoms residing in the triplet state. 

However, the collisions of bound electrons and free electrons will generally serve to relax the bound electrons to a distribution in statistical equilibrium with the electron temperature. Thus, it follows that the de-excitation rates, which have not been fully studied, must also play a role in re-equilibrating the bound electrons when the outflow cools due to adiabatic expansion. Therefore, we investigate the role of the de-excitation rates. Beyond, we note that due to the energetic proximity of $2^1$S and $2^1$P, electronic excitation should play a role, which we include in this work. Furthermore, we update the radiative decay rate for the states $2^1S \rightarrow 1^1S$, using the latest value from the \textsc{CHIANTI} database of $A_{2^1S\rightarrow 1^1S}=50.94~{\rm s}^{-1}$, which is significantly higher than that used by \citep{allan2024} of $A_{2^1S\rightarrow 1^1S}=51.3\times 10^{-4} {\rm s}^{-1}$. Beyond that, we note that the version of Penning ionization used in \citep{allan2024} is different in its ionization outcome from that in \citep{Oklopcic2018} in that the correct version produces $\rm H^{+} + e^{-}$ instead of $\rm H_0$. This change barely affects the physics of the lower atmosphere however, in which the Penning ionization functions to eliminate excess $\rm He 2^3 S$ in the presence of $\rm H^{+}$.
Collision rates between neutrals, ions and electrons are taken from \citep{Schunk2009} as analytic formulae.

\subsection{Methods of computation}

\label{sec:methods}
We use the \textsc{Chianti} package (version 9.0) and its associated database \citep{dere2019} to extract a complete set of electronic excitation and deexcitation rates. \textsc{Chianti} allows the user to systematically sift through all radiative, electron-induced state transition rates. Those rates $r$ are generally of a nontrivial form \citep{Oklopcic2018, delzanna2018}, containing complex state integrals. However, for inclusion in our radiation-hydrodynamic simulations we find that over the temperature range of interest for planetary atmospheric applications ($T\in[500:10^5]$K) we can fit the rates into a simple Arrhenius form $r=A\times T^{B} \times \exp(-C/T)$, with $T$ being the temperature normalized to 1 Kelvin, and the coefficients $A$,$B$,$C$ given in Table \ref{tab:all_rates}, and are plotted in Appendix \ref{sec:appendix_exdexfits}. These fits are accurate to 30\% or better over the range of interest in the simulations \ms{with the edges of the temperature range presenting the worst fit values.}

We update the photo- and thermochemistry module in  \texttt{aiolos} (described in Appendix \ref{sec:appendix_chemsolver}) to compute the level populations in neutral helium. 
This module solves rate equations for each species of the photoionizing form $\partial n_s/\partial t \propto \Gamma\; n_s$ or of the thermochemical form $\partial n_s/\partial t \propto r\times n_s\times n_{ss}  \times..$, where $r$ and $\Gamma$ need not be functions of the number densities $n_s$. Using the latter property and the fact that the rate transitions behave sufficiently close to their Arrhenius form, we treat the helium state transitions like thermochemical reactions. For the temperature, we adopt a minimum floor temperature of 500~K in our simulations. We also update \texttt{aiolos}'s multi-species solution method with a drift-adjustment scheme, which is described in Appendix~\ref{sec:appendix_drift}.
Lastly, we note the existence of problematic temperature spikes at the position of the multiple ionization fronts. Those arise due to the changes in mean molecular weight and temperature at the fronts, rendering the pressure discontinuous, ultimately cutting off the flow of tracers with the lower atmosphere. In order to smooth out those temperature spikes, we employ a simple causality-preserving temperature diffusion scheme, as presented in \cite{narayan1992}. Our numerical temperature diffusivity is low enough to not affect the outcome of our simulations significantly.

%

\begin{table*}
    \centering
    \begin{tabular}{c c c c c c c c }
    \hline
    $r^{\uparrow}_{ij}$ & A & B & C & $r^{\downarrow}_{ij}$ & A & B & C\\
    \hline
       $r^{\uparrow}_{12}$  & $7.79\times10^{-7}$ & $-0.531$ & $2.31\times10^5$ & $r^{\downarrow}_{12}$ & $2.60\times10^{-7}$ & $-0.531$ & $5.43\times10^2$ \\
       $r^{\uparrow}_{23}$  & $3.78\times10^{-5}$ & $-0.690$ & $1.02\times10^4$ & $r^{\downarrow}_{23}$ & $1.13\times10^{-4}$ & $-0.690$ & $9.16\times10^2$ \\
       $r^{\uparrow}_{27}$  & $1.52\times10^{-6}$ & $-0.438$ & $1.69\times10^4$ & ($r^{\downarrow}_{27}$) & $1.52\times10^{-6}$ & $-0.438$ & $7.03\times10^2$ \\
       $r^{\uparrow}_{37}$  & $1.61\times10^{-7}$ & $0.249$ & $7.21\times10^3$ & ($r^{\downarrow}_{37}$) & $5.36\times10^{-8}$ & $0.249$ & $2.25\times10^2$ \\
       \hline
    \end{tabular}
    \caption{Table for our fit coefficients of $r^{\uparrow}_{ij}$ and $r^{\downarrow}_{ij}$. The $2^1$P (level 7) is severely underpopulated compared to the other levels due to its high radiative de-excitation coefficient towards the ground state $A_{2^1P \rightarrow 1^1S}=1.8\times10^{9}s^{-1}$, hence we have bracketed out de-excitation rates from that level, emphasizing that it can be neglected in modelling, but still list the rates for the sake of completeness. The radiative de-excitation rates from \citep{allan2024} and the Chianti database are $\rm A_{2^1S\rightarrow\, 1^1S,\;old}=51.3\times10^{-4}s^{-1}$ and $\rm A_{2^1S\rightarrow\, 1^1S,\;new}=50.94\,s^{-1}$, respectively. }
    \label{tab:all_rates}
\end{table*}

\begin{figure}
\centering
	\includegraphics[width=0.8\columnwidth]{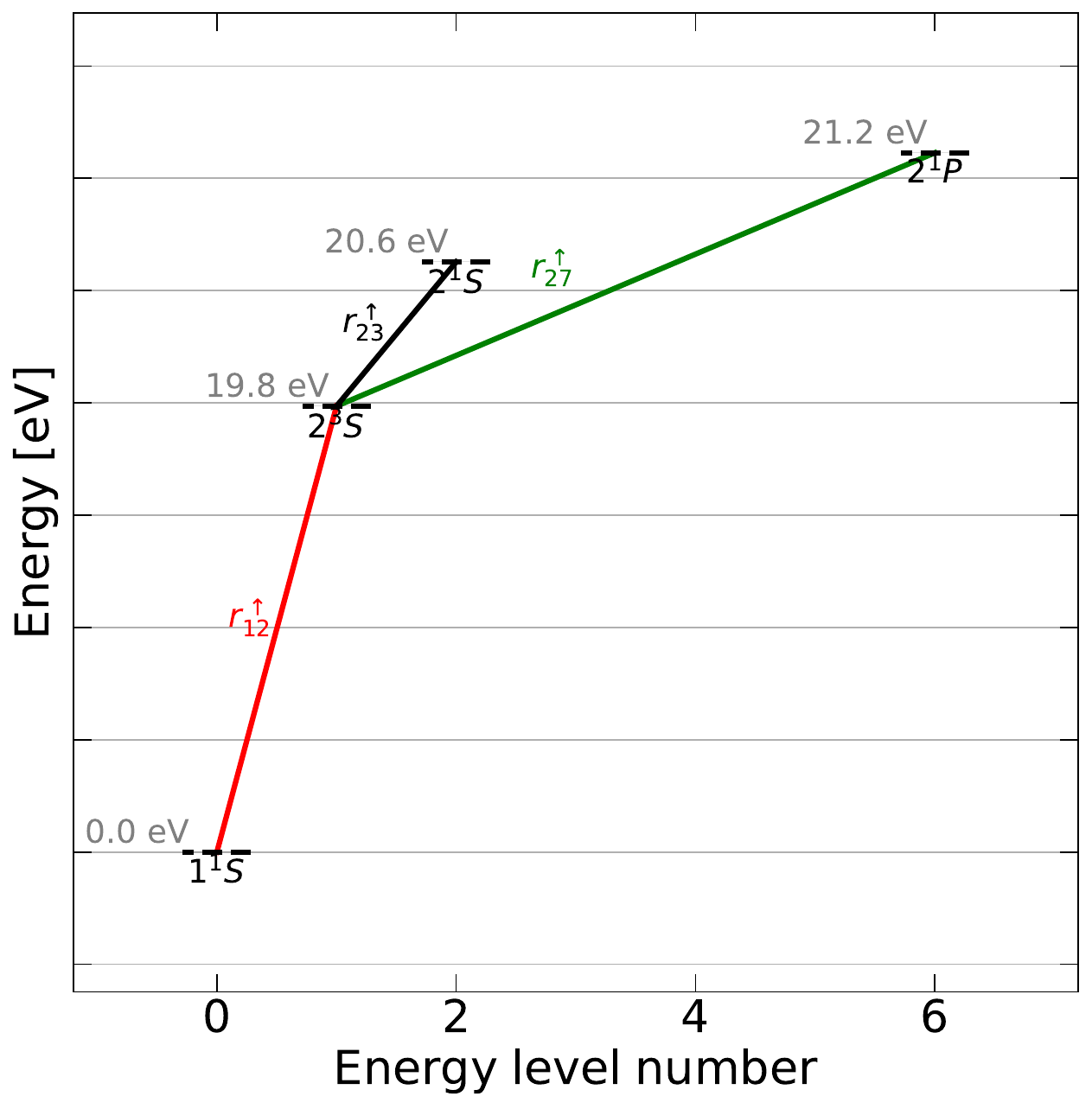}
     \includegraphics[width=0.8\columnwidth]{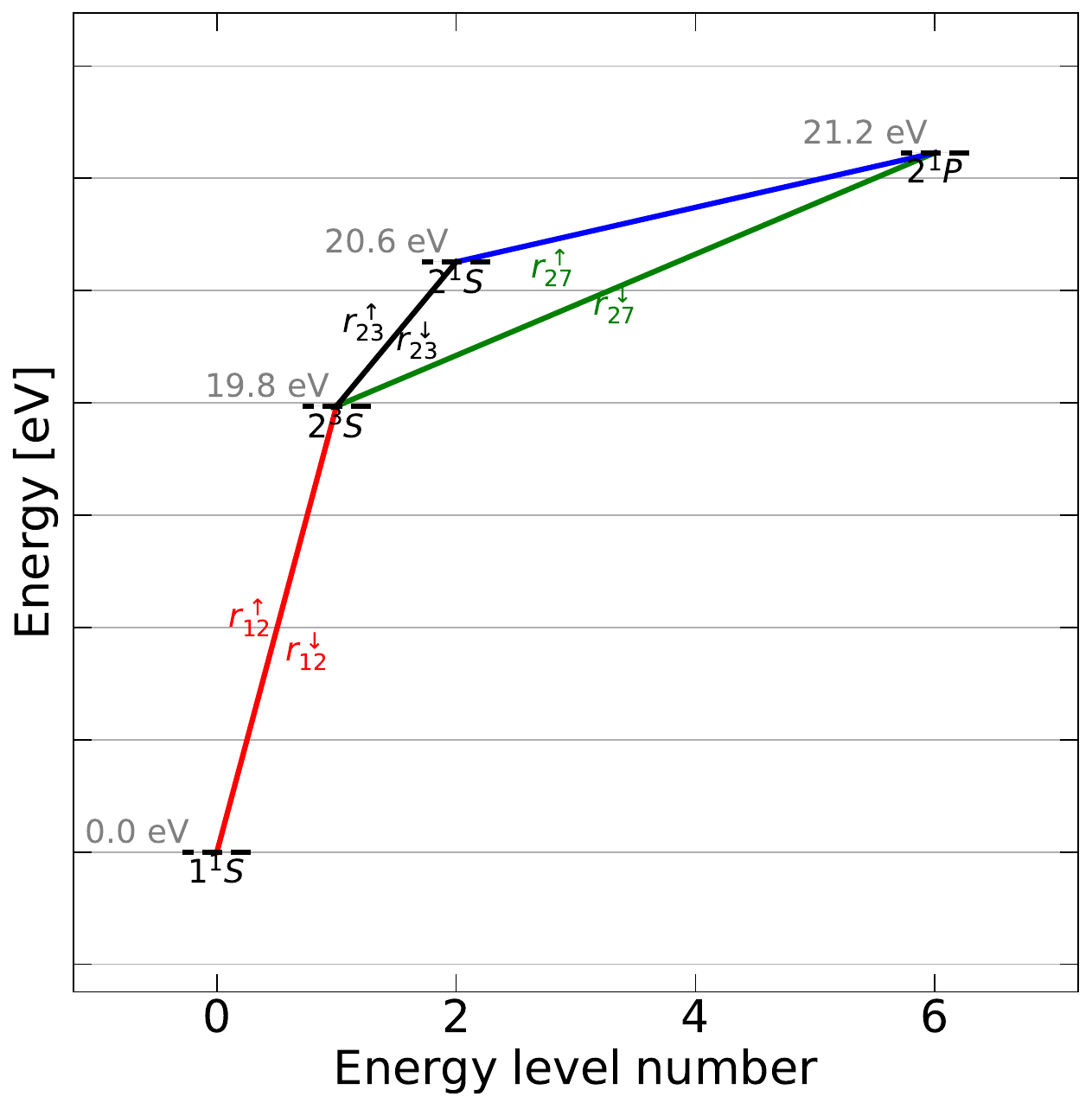}
    \caption{Energy level structure and active transition rates for two different temperatures, 10,000~K (Top) and 1,000~K (Bottom). Newly added rates are explicitly marked in Table \ref{tab:all_rates}. Note the additional pathways that become important for the triplet state at low temperatures. }
    \label{fig:energylevels}
\end{figure}

\begin{figure*}
\centering
 	\includegraphics[width=0.495\textwidth]{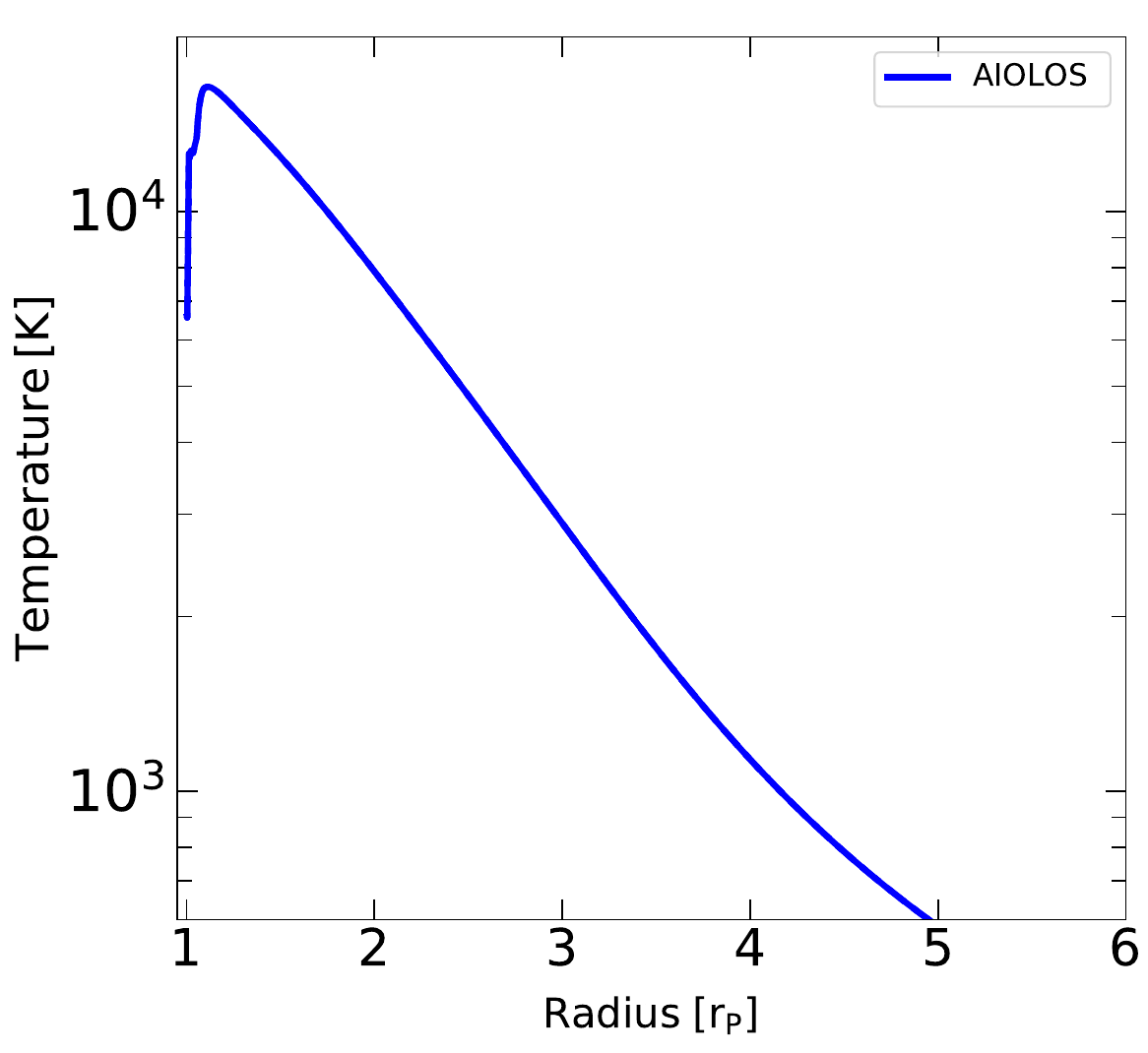}
    \includegraphics[width=0.495\textwidth]{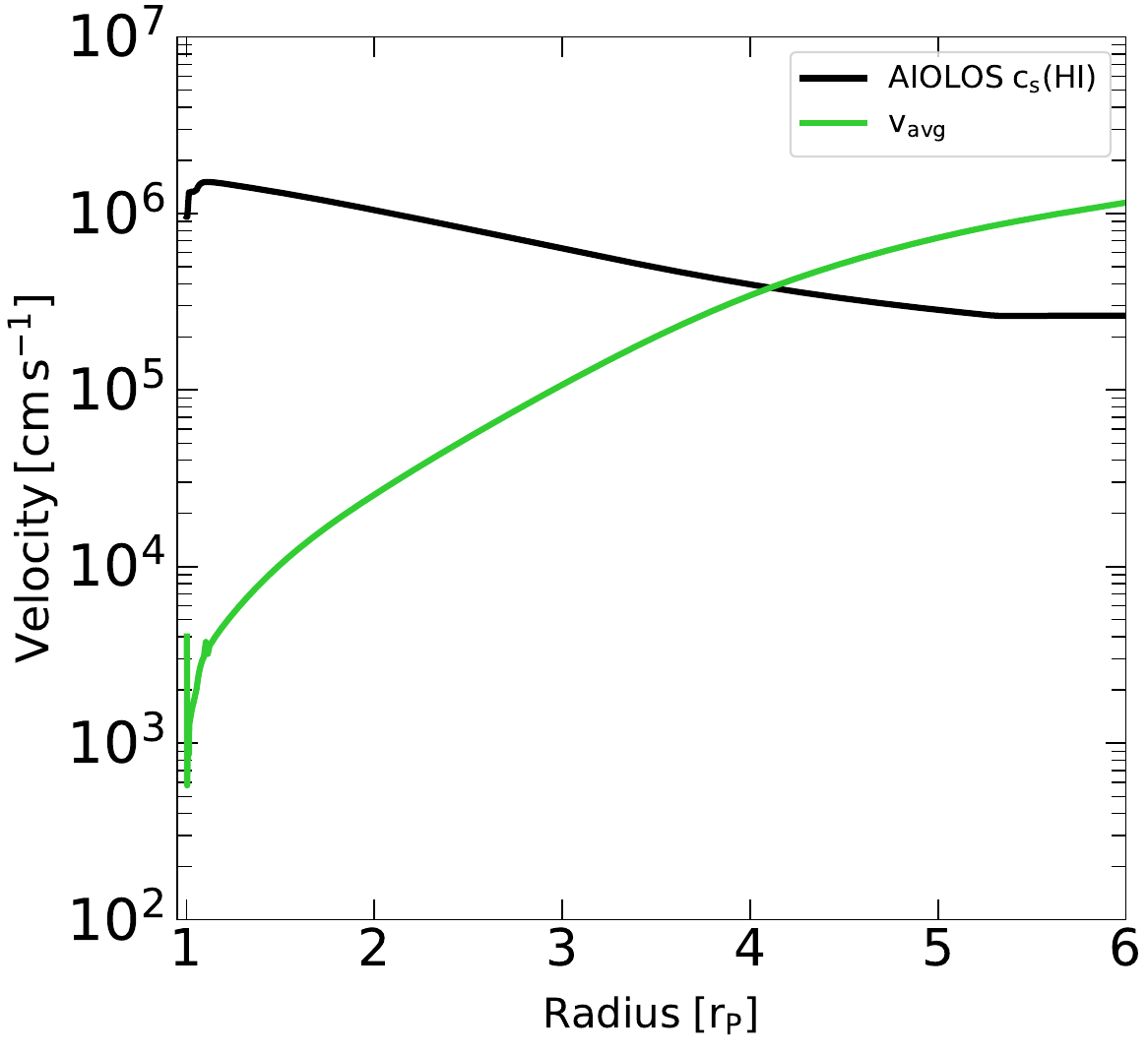}
        
    \includegraphics[width=0.495\textwidth]{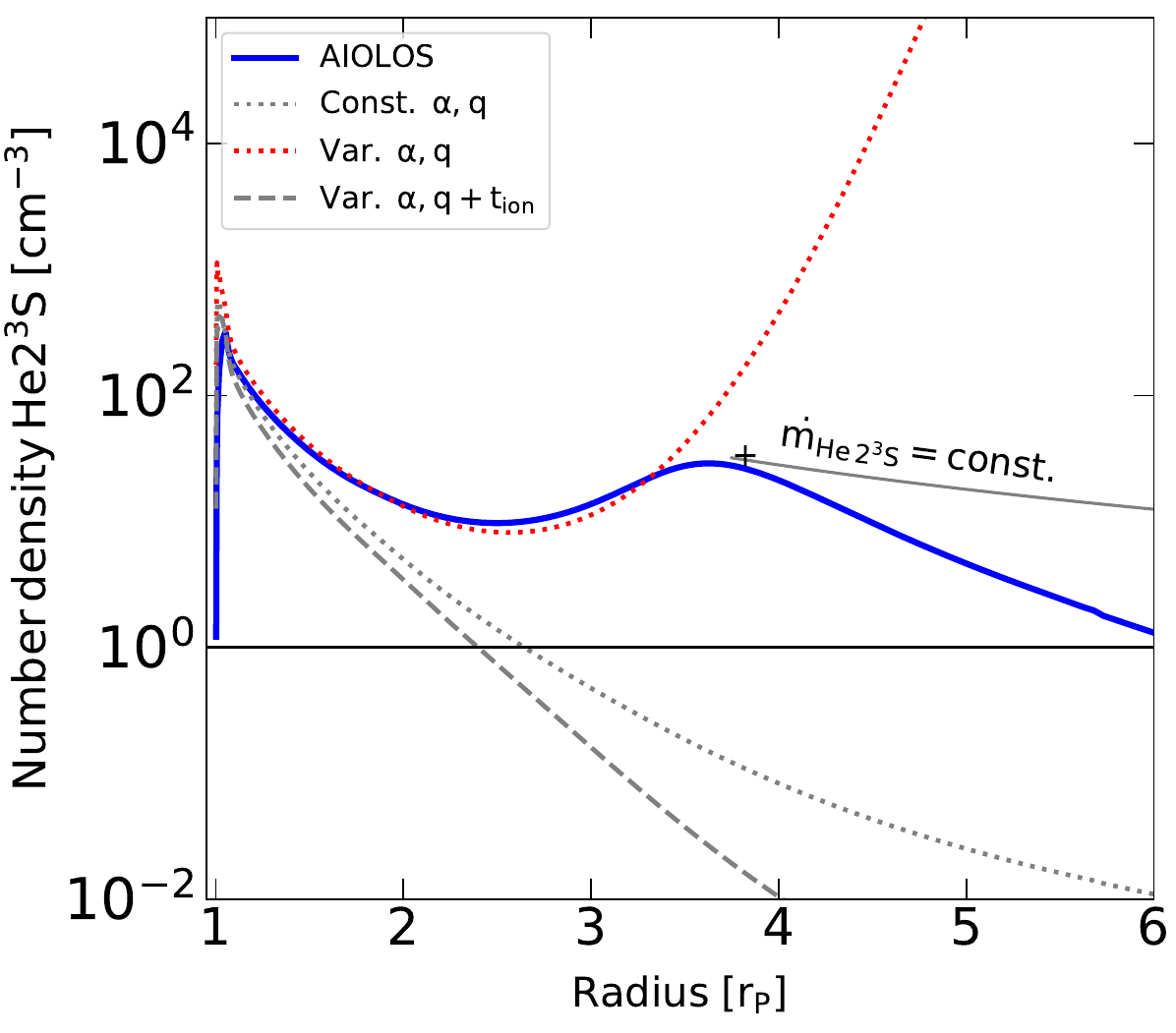}
  	   \includegraphics[width=0.495\textwidth]{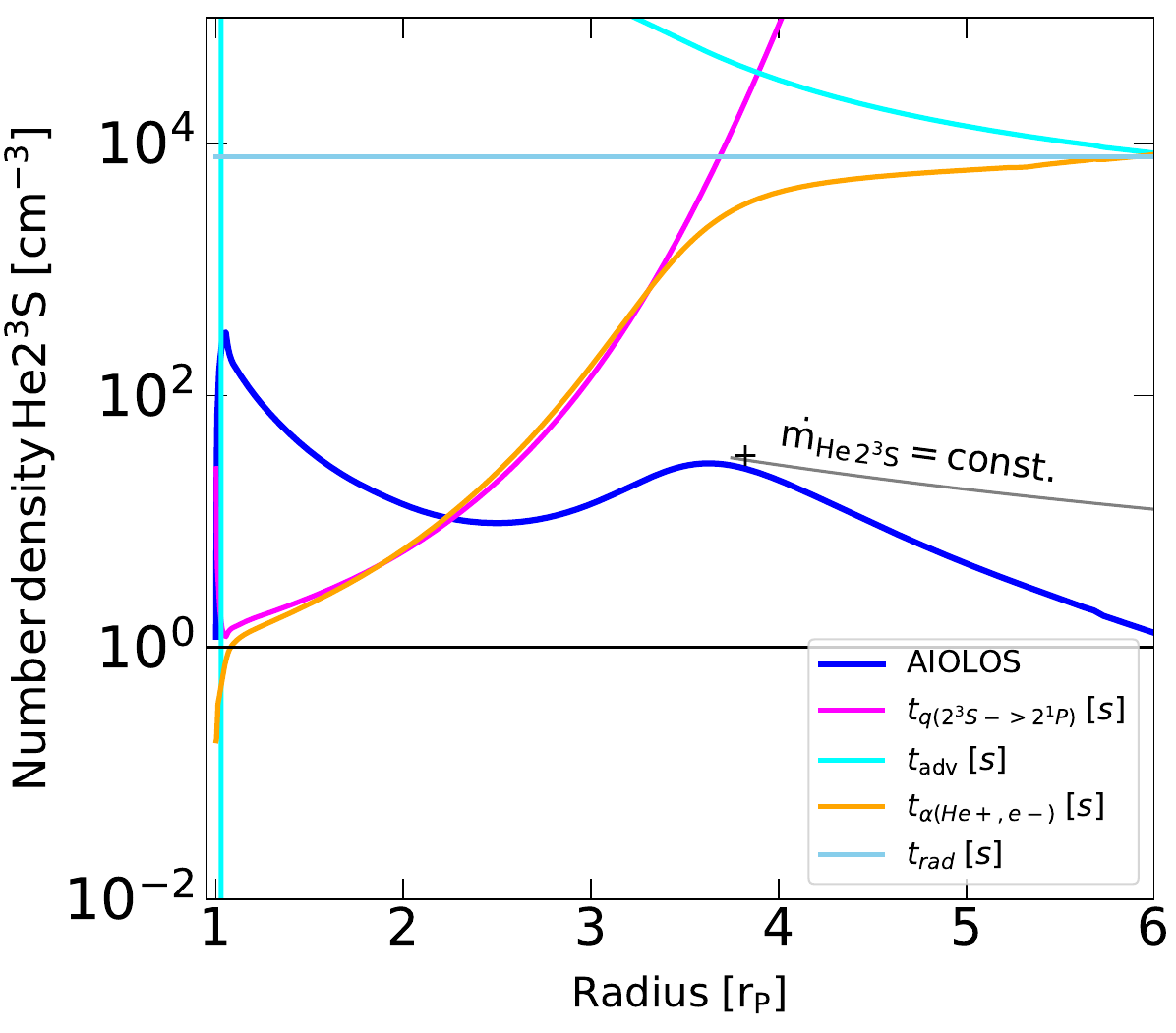}
 
    \caption{A sketch of the physics of the Helium population bump based on our simulated data: While the high temperatures near the planet (top left) accelerate the flow to the speed of sound (top right), driving vigorous mass-loss, a maximum in the triplet population can be produced due to the adiabatic cooling of the gas as it expands into space.
    The comparison of simulation data with simple approximations of Eqn. \ref{eq:triplet_rate_solution_simple} and timescales allows a deeper understanding of the physics reflected in the shape of the triplet population curve (bottom): On the left we plot population curves computed for the coefficients $\alpha,\;q$ evaluated at a constant temperature of $10^4$K (const. q curve), when the coefficients follow the local temperature (variable q) and when ionization is taken into account. 
    For this example, the simulation ran without the triplet ionization term. On the right, we show that the bump is produced by a switch of the balance of recombination with collisional excitation timescale to the balance of recombination to radiative, advective (and eventually ionization) timescales. The undershoot of the mass-loss curve  $\rm \dot{m}_{\rm He2^3S}=const.$ emphasizes that past the bump, additional loss processes beyond advection are at play.}
    \label{fig:physics_intro_to_triplet}
\end{figure*}

\section{Helium level populations}
\label{sec:heliummodels}

Since outflows from high surface gravity planets cool significantly as they expand \citep[e.g.][]{MurrayClay2009,owenalvarez2016}, we discuss three main pathways that change the $\rm He 2^3S$ population at low temperatures, two of them new. A  depiction of population pathways for the neutral helium energy levels for a typical hot Jupiter is shown in Fig. \ref{fig:energylevels}, demonstrating that additional pathways become important for the triplet state at low temperatures. 

\subsection{General physics of the He$2^3$S profiles}

In general, the dominant population pathway of the He$2^3$S state is recombination from He$^{+}$ ions \citep[e.g.][]{osterbrock2006,Oklopcic2019}. The dominant sink terms are collisional excitation towards the $2^1$S state and ionization back into the He$^{+}$ state; then, dropping second order processes in order to obtain a simple quantitative comparison population curve, can an approximation to the He$2^3$S state population be found by solving the approximate rate equation \citep[e.g.][]{Oklopcic2019,biassoni2024}:
\begin{align}
    \frac{\partial n_{2^3S}}{\partial t} = + \alpha(T_e) n_{e} n_{He^{+}} - q(T_e) n_{e} n_{2^3S} - t_{\rm ion} n_{2^3S}
    \label{eq:triplet_rate_equation_simple}
\end{align}
where $n_i$ are the number densities of the different states, $\alpha$ is the collisional case-B recombination rate, $q$ is the excitation rate of bound helium electrons due to collisions with free electrons at a temperature $T_e$ and $t^{-1}_{\rm ion} = \int d\nu \,F/{h\nu} \times \kappa$ is the optically thin ionization time computed from the photon flux $F$, photon energy $h\nu$ and the absorption cross-section $\kappa$ for all wavelengths. We intentionally omit several additional terms in this rate equation \citep[see, e.g.][]{Oklopcic2018, allan2024, ballabio2025} for the sake of simplicity, but our simulations include all the terms common in the literature and the ones which we add in the subsequent sections.
Eqn. \ref{eq:triplet_rate_equation_simple} has a steady-state solution
\begin{align}
    n_{2^3S}(T_e) = \frac{\alpha(T_e)}{q(T_e) + t_{\rm ion}/n_e} n_{He^{+}}
    \label{eq:triplet_rate_solution_simple}
\end{align}
which we will use as a reference state for the He$2^3$S populations.
While \texttt{aiolos} can handle separate temperatures for all species, due to the high-density outflows for hot Jupiters, in most of this work, we keep all temperatures for the species ($s$) at the same value representing the average temperature $T=T_{s}$, where we investigate the effects of $T_{e}\neq T$).

We now briefly discuss the physics shaping the $n_{2^3S}(T_e)$ population as a function of distance from the giant planet in its outflow. 
As shown in Fig. \ref{fig:physics_intro_to_triplet}, the temperature in the outflow drops dramatically from $\simeq10^4$K to $\simeq 10^3$K due to adiabatic cooling. 

Hence, while the triplet population initially follows Eqn. \ref{eq:triplet_rate_solution_simple} for a $n_{2^3S}(T=10,000K)$ power-law curve ("const q"), a change in slope in the population occurs 
as the dominant triplet loss rate $q(T_e)$ drops significantly at lower temperatures ("variable q"). This effect has already been pointed out in \cite{yan2022} and \cite{biassoni2024} as the cause of the population bump.

\begin{figure*}
 \includegraphics[width=0.47\textwidth]{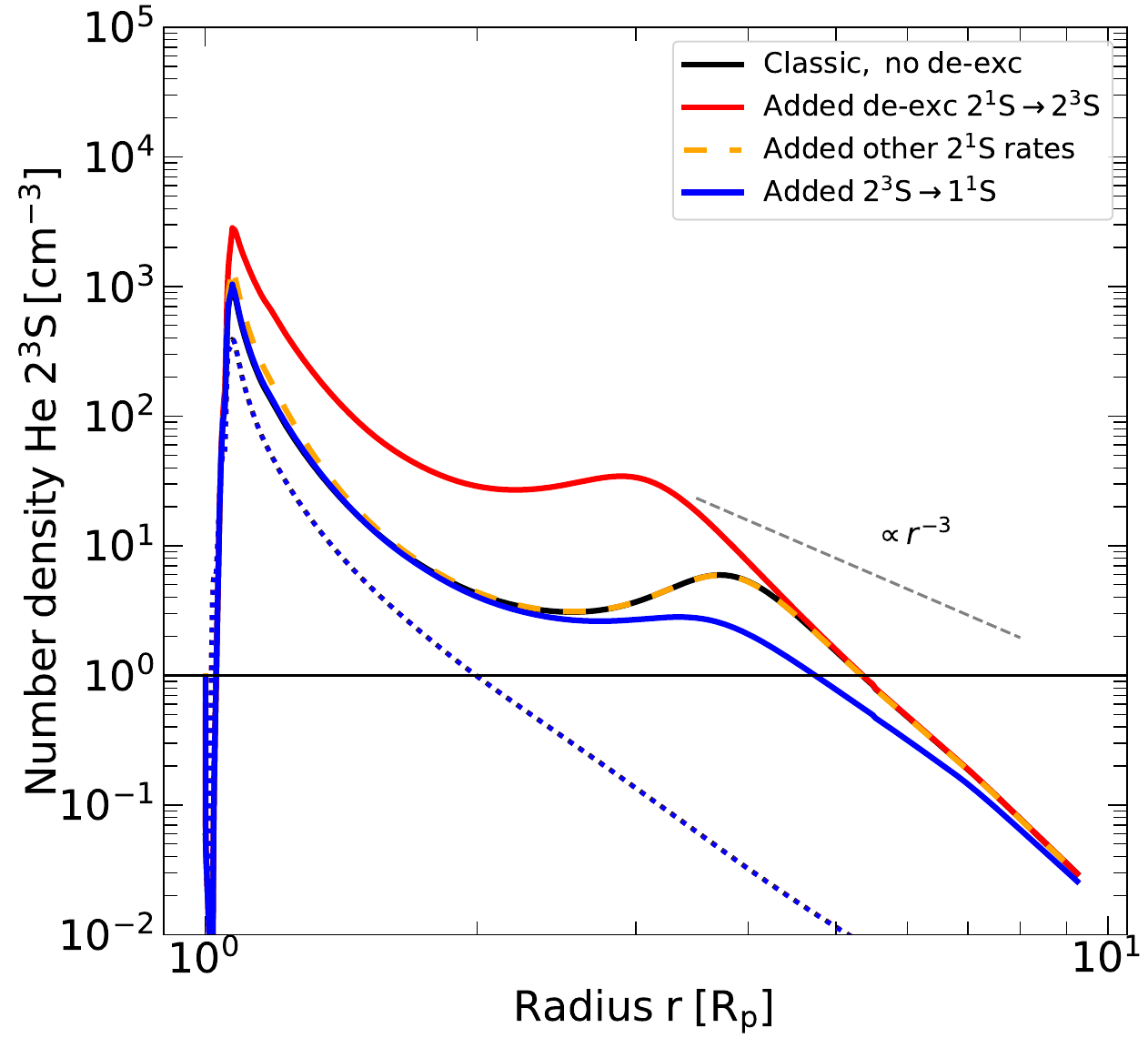}
 \includegraphics[width=0.47\textwidth]{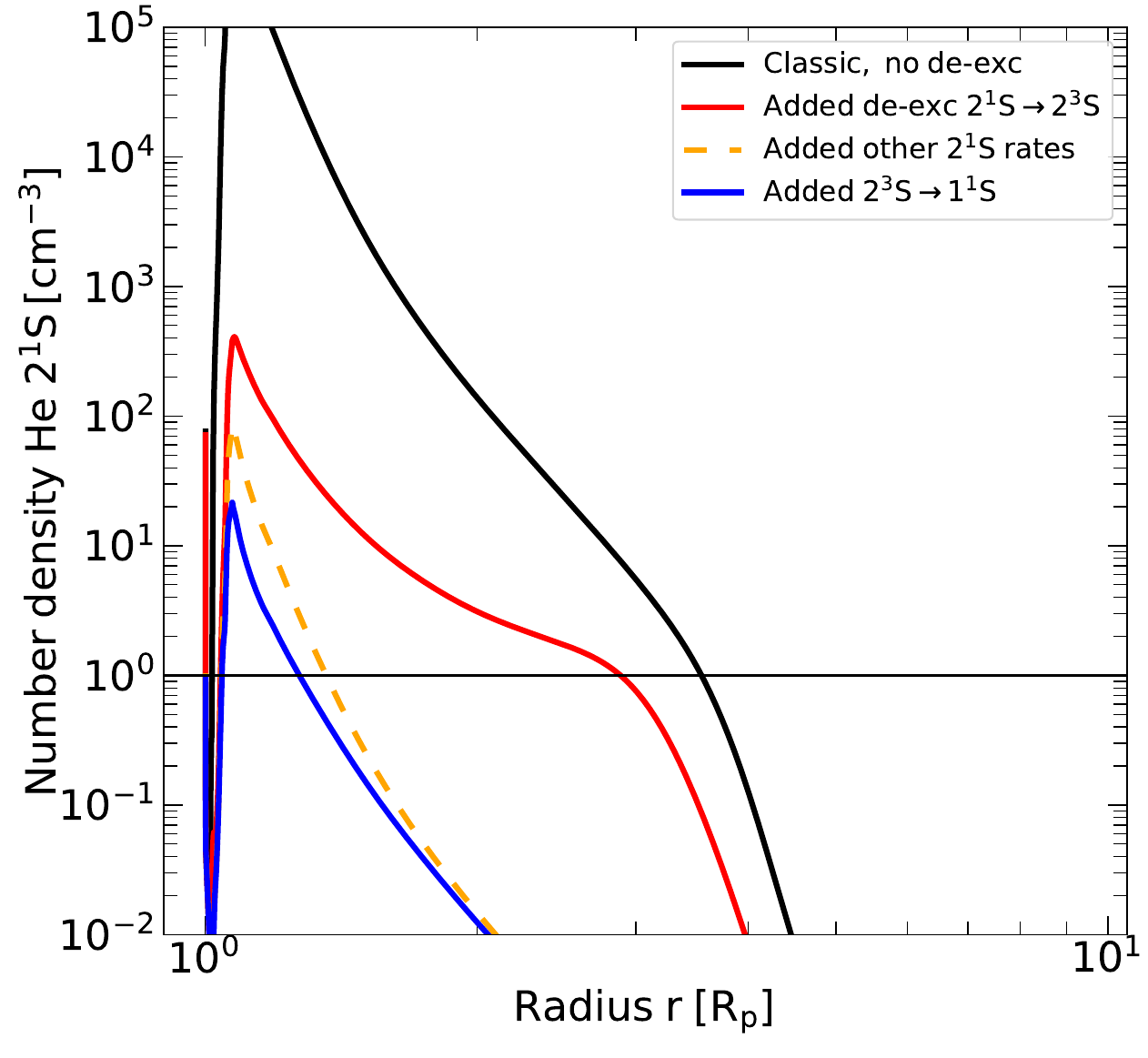}
    \caption{ The metastable triplet population profiles, with varying inclusion of the new rates (left). Populations in the $2^1$S states (right) are added to aid in understanding why the black and orange lines overlap. The dashed grey line indicates a constant contribution to the optical depth per radial interval, indicating that the transit signal resulting from such a hypothetical population will be dominated by contributions at a few planetary radii. As before, the dotted grey line indicates the isothermal population level.}
    \label{fig:effect_of_rates_on_triplet}
\end{figure*}

The radial increase in population levels continues until the flow advection timescale becomes fast enough to compete with the excitation to $2^1$S - the triplet population level effectively becomes ``quenched'' at this point, marked as "+" in Fig. \ref{fig:effect_of_rates_on_triplet}. The quenched flow, however, does not escape with a constant triplet fraction (solid grey curve), i.e.: 
\begin{align}
    \dot{m}_{2^3S} = 4\pi\, r^2\, v\, n_{2^3S}\, m_{He}
\end{align}
due to ionization and the radiative decay $\rm 2^3S \rightarrow 1^1S$ consistently removing triplet electrons. We note that the relative importance of advection compared to the other processes will be a function of the spectral type of the host star, adding complexity to previous analyses (see Appendix A.2 in \citep{biassoni2024}).


We note that for $T \geq 5\times10^3$K, corresponding to $r \leq 2.5 r_p$, the nonconstant rates, as well as the new rates we add, see Fig. \ref{fig:effect_of_rates_on_triplet}, do not play a major role, consistent with the simplified scheme of \citet{Oklopcic2018}.
We propose to use the population bump, occurring in He$2^3$S occurring at cold temperatures, as a signpost of adiabatic cooling in outflows. Since adiabatic cooling requires the gas to maintain a thermal statistical equilibrium through mutual collisions, detecting adiabatic cooling would confirm the outflow is a collisional hydrodynamic outflow on larger scales than typically probed by metal transits. Thus, one could confirm the outflow is collisional on a scale up to the sonic point.  Since the local maximum is connected to the local velocity structure, it is possible that this enhanced opacity would be located sufficiently far into the line wings to be detectable in high-resolution line profiles using ground-based spectrographs.

We also aim to develop a simple and analytical understanding of whether to expect a finite transit signal from our triplet populations or unbound transit solutions akin to the detected giant tails in \citep{Zhang2023}. For this, we compare the population profiles against a $r^{-3}$ power law in Fig. \ref{fig:effect_of_rates_on_triplet}. Such a population profile would correspond to constant optical depth contributions per unit radius $dr$. As the populations we find drop steeper than $r^{-3}$ at large radii, the transit signal will be dominated by the contribution from inside the planet's Hill sphere. 

\subsection{Why de-excitation from He $2^1$S is not a source of large discrepancies}

We now discuss why the population in the $2^1$S, somewhat surprisingly, cannot repopulate the $2^3$S level. \citet{allan2024} found a significant build-up of the $2^1$S population, which we reproduce in Fig. \ref{fig:effect_of_rates_on_triplet} (black lines, both panels). Adding the de-excitation process from $2^1$S$\rightarrow2^3$S, with the corresponding de-excitation rate $r^{\downarrow}_{32}$, results in repopulating $2^3$S (red lines). This occurs already when the gas temperature is $10^4$K and creates an even stronger excess at lower temperatures of more than a factor of $10$ above the initial computation.
However, the action of the other new processes, particularly the electron excitation $2^1$S$\rightarrow 2^1$P and the updated radiative decay rate $\rm A_{2^1S\rightarrow 1^1S}$, result in strong depletion of the $2^1$S-state. Thus, the now highly depleted $2^1$S-state is no longer available to repopulate $2^3$S (orange dashed lines).
Since the state $2^1$P is always strongly depopulated due to its rapid radiative decay rate towards the ground state, $2^1$S$\rightarrow 2^1$P can be effectively understood and modelled as an indirect ground-state recombination $2^1$S$\rightarrow 1^1$S, similar to the original reduced model of \citep{Oklopcic2018}. 

We ran simulations without including the $2^1$S state by sending all excited electrons from $2^3$S directly to $1^1$S. In three setups of an HD189733b-equivalent being hosted by a G2, K2 and M2 star, we found this approximate approach produced negligible differences to the simulations which included the $2^1$S state. We note that the choice of the radiative rate $A_{2^1S\rightarrow 1^1S}$ also plays a role in setting the final populations - using the earlier slower value used in \citep{allan2024}, results in incomplete evacuation of the $2^1$S state and partial refilling of $2^3$S, where the inclusion of the $2^1$S state is required. 

\begin{figure}
 \includegraphics[width=0.9\columnwidth]{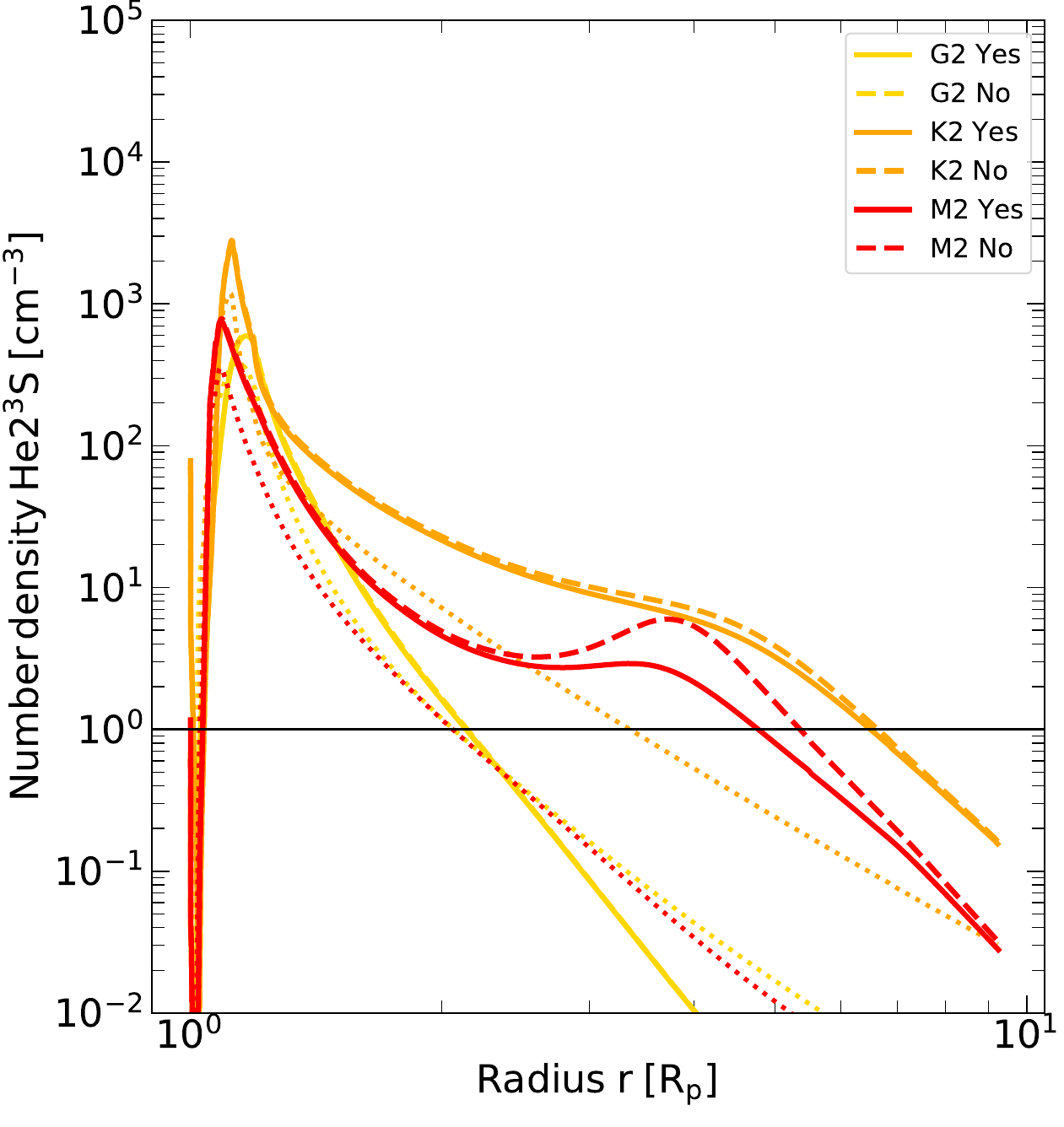}
  \includegraphics[width=0.9\columnwidth]{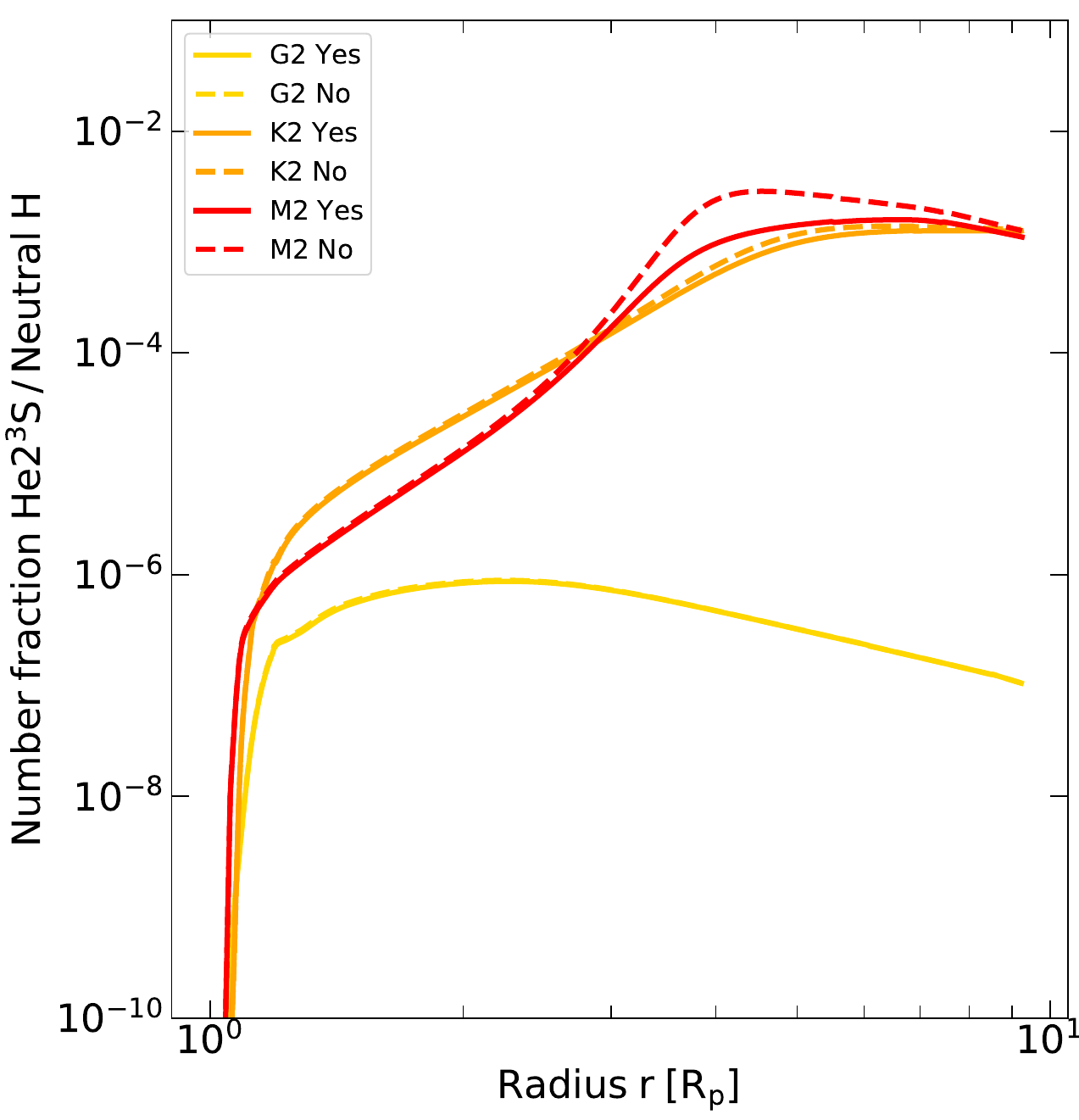}
    \caption{ The stellar dependence of the helium triplet number density (top) and its fractional population (bottom). The solid lines include the remaining important rate, $2^3$S $\rightarrow$ $1^1$S, while the dashed lines do not. Note the the swap of relative ordering in density compared to the fraction between the orange and red curves. Reference isothermal solutions are dotted and computed from Eqn. \ref{eq:triplet_rate_solution_simple}. }
    \label{fig:effect_of_stars_on_triplet}
\end{figure}

\subsection{De-excitation to the ground state and dependence on stellar types}

We now discuss another important de-excitation pathway: $\rm 2^3$S $\rightarrow$ $\rm 1^1$S, which can significantly suppress the formation of the $\rm He 2^3$S population bump (blue lines in Fig. \ref{fig:effect_of_rates_on_triplet}).
This leaves a weaker excess over the isothermal model and undershoots it at the coldest temperatures.

However, this process competes with photoionization, which for early-type stars can typically be the dominant depopulation process for $2^3$S \citep{Oklopcic2019,allan2024, biassoni2024}, due to the large amount of available NUV photons, and the photoionization edge of the $2^3$S state reaching $\rm 4.67 eV$ \citep{norcross1971}. Hence, we investigate the importance of this de-excitation process as a function of stellar type, the results of which can be seen in Fig. \ref{fig:effect_of_stars_on_triplet}, where we compare the population curves with and without the $\rm 2^3$S $\rightarrow$ $\rm 1^1$S process. As before, the investigated planet is a HD189733b analogue, orbiting a G2, K2 and M2 dwarf at a distance of $0.03$ AU each.

It becomes evident that there is no difference in populations for the G2 star, a difference of about 10$\%$ for the K2 star and a 100$\%$ difference for the M2 dwarf.
Qualitatively, this is fully in line with expectations from the intensity of NUV radiation which each star emits.

It is further interesting to note that, particularly for M-dwarfs, there is a stark contrast between the expected transit depths (predicted to have the deepest transits of all stellar types, reaching $\geq 10 \%$, \citealt{biassoni2024}) and the dearth of detections for this stellar type \citep{dosSantosreview}. Thus, while the electron-induced de-excitation to the ground state moves the theoretical predictions into the direction observations suggest, the effect we see in our simulations is insufficient to fully explain the non-detections.

\begin{table*}
\hspace{-1.0 cm}
    \centering
    \begin{tabular}{c | c | c | c | c | c}
    Prototype\\name & Star used & Refererence & $\dot{m}_{H}$ & $\dot{m}_{He}$ & Ratio \\
    \hline 
     G2 & Sol, contemporary & \citep{claire2012} &3.6e+10 & 9.2e+09 & 0.26 \\
     K2 & HD189733 & \citep{bourrier2020} & 9.02e+09 & 1.82e+09 & 0.20 \\
     K2 & HD189733 & \citep{lampon2021} & 4.41e+11 & 9.05e+10 & 0.21 \\
     M2 & GJ3470 & \citep{lampon2021} & 1.07e+10 & 2.57e+09 & 0.24 \\
     HD209458 & Sol, contemporary & \citep{claire2012} & 9.44e+10 & 2.53e+10 & 0.27 \\
    \end{tabular}
    \caption{List of stellar types used, source of the spectra and mass-loss rates for HD189733b (all except for the last entry) and HD209458b analogues orbiting said stars. The ratio of mass-loss rates indicating fractionation (values lower than 27$\%$) are computed with the full numerical precision of the mass-loss rates. }
    \label{tab:stellartypes}
\end{table*}

\begin{figure}
    \centering
    \includegraphics[width=\linewidth]{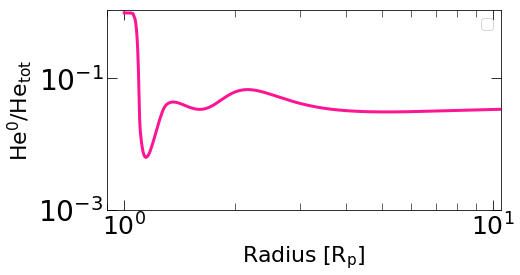}
        
    \includegraphics[width=\linewidth]{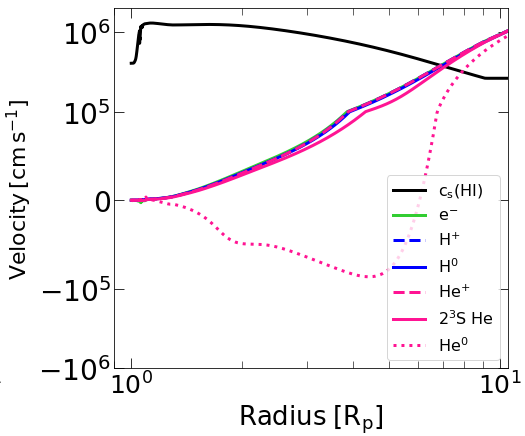}
    \caption{While Fig. \ref{fig:physics_intro_to_triplet} showed mean velocities for all species, here we plot the velocities for individual species. Fractionation between ions and neutrals explain different species' velocities. It is evident that the fractionation behaviour can differ dramatically for triplet and ground-state helium, which can even develop negative velocities. Triplet helium is significantly less fractionated as it inherits its velocity from ionized helium through recombination.  }
    \label{fig:velocity_fractionation_intro}
\end{figure}

\subsection{Fractionation between different species of Helium}

In a multi-species gas, each species' individual velocity does not have to be the same. Generally, the velocity profiles are controlled by gravity and pressure gradients in addition to collisions between the various species. Further modification of the momentum is possible when species transfer between each other through photochemical processes. This final process is particularly relevant for the helium triplet.
{A way to quantify the amount of fractionation of a heavy species $s'$ in an outflow driven by a light species $s$ is to determine the ratio of their mass fluxes 
} 
\begin{align}
    \frac{\dot{m}_{s'}}{\dot{m}_s}(r) = \frac{\rho_{s'}v_{s'}}{ \rho_s v_s}(r)
\end{align}

{evaluated at the sonic radius $r_s$. This value will be identical to the base value of density ratios, $27\%$ in our case, in a well-coupled non-fractionating outflow of hydrogen and helium. It will be lower in fractionating outflows where $s'$ is left behind relative to $s$. We note that for identical velocities the mass flux ratio is also the ratio of atoms in the upper atmosphere. We omit the normalization to the atmospheric base density $\rho_{s'}/\rho_s(0)=27\%$ \citep{zahnle1986, hunten1987} for reasons of simplicity. }\\

Consider He$1^1$S initially at rest, and $\rm He^{+}$ outflowing while entrained in a planetary wind. Due to ionization, there will be constant upcycling of slow, neutral He into $\rm He^{+}$. The latter is subsequently accelerated by drag with protons and electrons. 
As a number of slow atoms is added to the fast moving ions, the summed momenta will effectively slow down the $\rm He^{+}$ flow through the ionization process. The opposite is the case for electronic recombination: fast $\rm He^{+}$ particles are added to the slow He$1^1$S reservoir, accelerating it. 
The same effect acts on the excited neutral He$2^3$S, but its entire population is quickly recycled into other species, before its constituent atoms can de-accelerate. Therefore what appears as rapidly outflowing steady-state population, is actually a population in a dynamic exchange-equilibrium.


We stress that this ``inheritance'' of momentum plays the role of an effective acceleration. This produces fast He$2^3$S, whereas classically \citep[e.g.][]{zahnle1986, hunten1987} one might expect all neutral Helium to possess slower velocity and fractionate. Details on the chemistry solver and the numerics of momentum transport between species are outlined in Appendix \ref{sec:appendix_chemsolver}.
Our simulation results for a HD209458b-like outflowing atmosphere in which helium species are dragged by hydrogen, protons and electrons are shown in Fig. \ref{fig:velocity_fractionation_intro}, where the velocity structures for all species are now shown. 
He$1^1$S is slow and fractionates at the base of the wind, i.e. it reaches a profile compared to the rest of the outflow, and even negative velocities, similar to what was seen in \citep{xing2023}. This species only reaches the sound speed because it remains collisional, and the tidal potential from the star allows it to escape. As Helium here is largely ionized, the average velocity, previously shown in Fig. \ref{fig:physics_intro_to_triplet}, is dominated by its contributions from $\rm p^{+}$ and $\rm He^{+}$. Thus, the He$2^3$S, having inherited its velocity from the ionized helium, remains fast close to the planet and traces the majority of helium in the outflow. This would be different in mostly neutral outflows, where although He$2^3$S remains fast, it does not correspond to the majority of escaping Helium mass flux and hence becomes unobservable (see the next section).

\begin{figure}

  \includegraphics[width=\columnwidth]{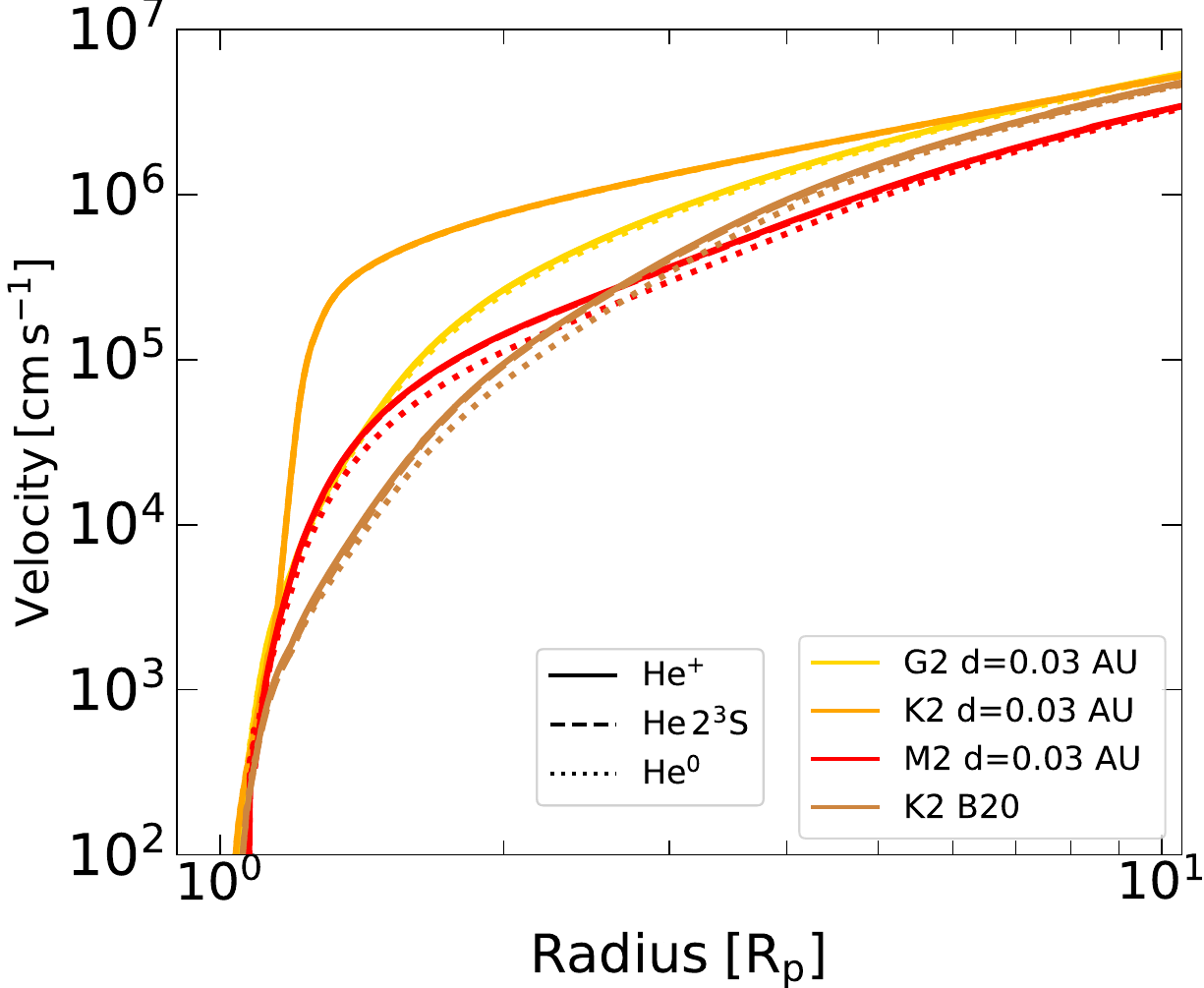}
  \includegraphics[width=\columnwidth]{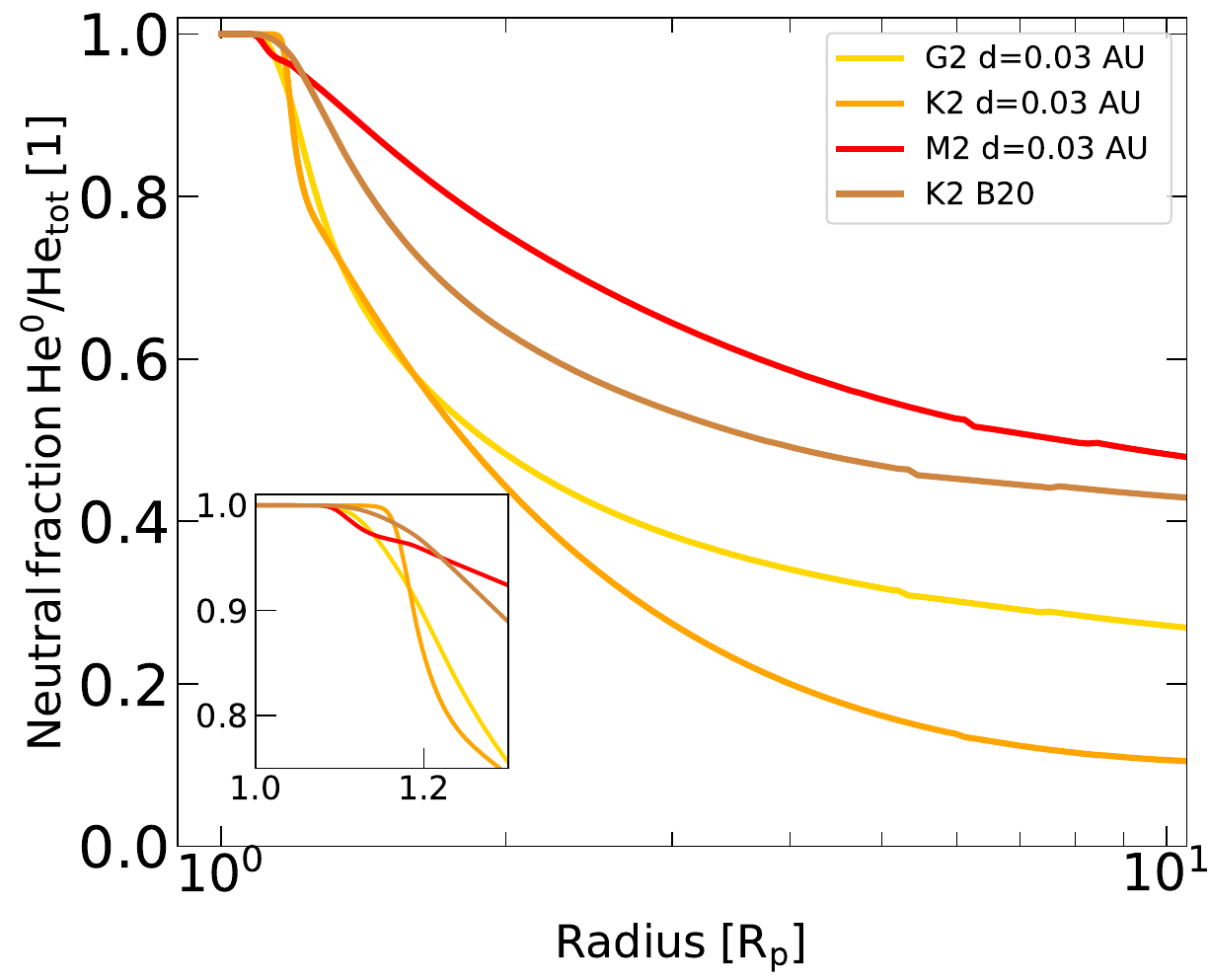}
    \caption{ The stellar dependence of fractionation, including the remaining important rate, $2^3$S $\rightarrow$ $1^1$S. The upper panel shows the velocity profile, while the lower panel shows the helium neutral fraction, the inset shows the neutral fraction at the base of the flow, which is crucial to determine the fractionation. These plots use the same simulation data as in Fig. \ref{fig:effect_of_stars_on_triplet}, with all rates included.}
    \label{fig:velocity_fractionation_stars}
\end{figure}

As the fractionation of helium might also carry consequences for the evolution of lower-mass planets \citep{Malsky2023}, we now investigate velocity structures species-by-species for analogues of HD 189733~b orbiting different stars in Fig. \ref{fig:velocity_fractionation_stars}. The corresponding mass-loss rates and mass-loss fractions $\dot{m}_{\rm H,tot}/\dot{m}_{\rm He,tot} $ are listed in Table \ref{tab:stellartypes}. Given the high gravity of HD 189733~b, those are likely to represent cases where fractionation is maximised. 

We irradiate the gas giant with a typical G2, M2 spectrum and two different literature alternatives for the K2-star, HD 189733.
We note that the spectrum by \citep{lampon2021} is a simulated spectrum and yields a hydrogen mass-loss rate of $\rm 4\times10^{11} g/s$ . The spectra from other approaches \citep[e.g. line based reconstruction techniques as in][]{salz2016, bourrier2020} can yield significantly differing mass-loss rates, typically around $\rm 1\times10^{10} g/s$ for HD189733b. We note that recently, \cite{zhang2024} reported discrepancies of similar nature and magnitude for the spectral modelling of TOI-836.

From these results, it becomes evident that the K2 spectra, although their mass-loss rates can be very high, trigger significantly stronger fractionation compared to the other stellar types. The reason for this lies in the deeper penetration of high-energy photons in the G2 and M2 spectra, (see the inset in Fig. \ref{fig:velocity_fractionation_stars}) ionizing Helium at lower radii. Because all species velocities are close to each other in each simulation, this shows that it is not only the hydrogen mass-loss rate that drives fractionation, but instead, the ionization state of the gas at the outflow base also plays a critical role in coupling $\dot{m}_{H}$ and $\dot{m}_{He}$ efficiently.
The overall effect of the ionization state on fractionation arises because Coulomb collisions between ionized species are much more efficient at momentum transport than neutral-neutral collisions \citep{schunk1980}. Higher ionization fractions can overcome lower densities to result in stronger collisional coupling. 

%

We now focus on the host stars with the largest discrepancy between observed and expected transit depths, and investigate what role fractionation would play in their outflows.

\begin{figure*}
 \includegraphics[width=0.32\textwidth]{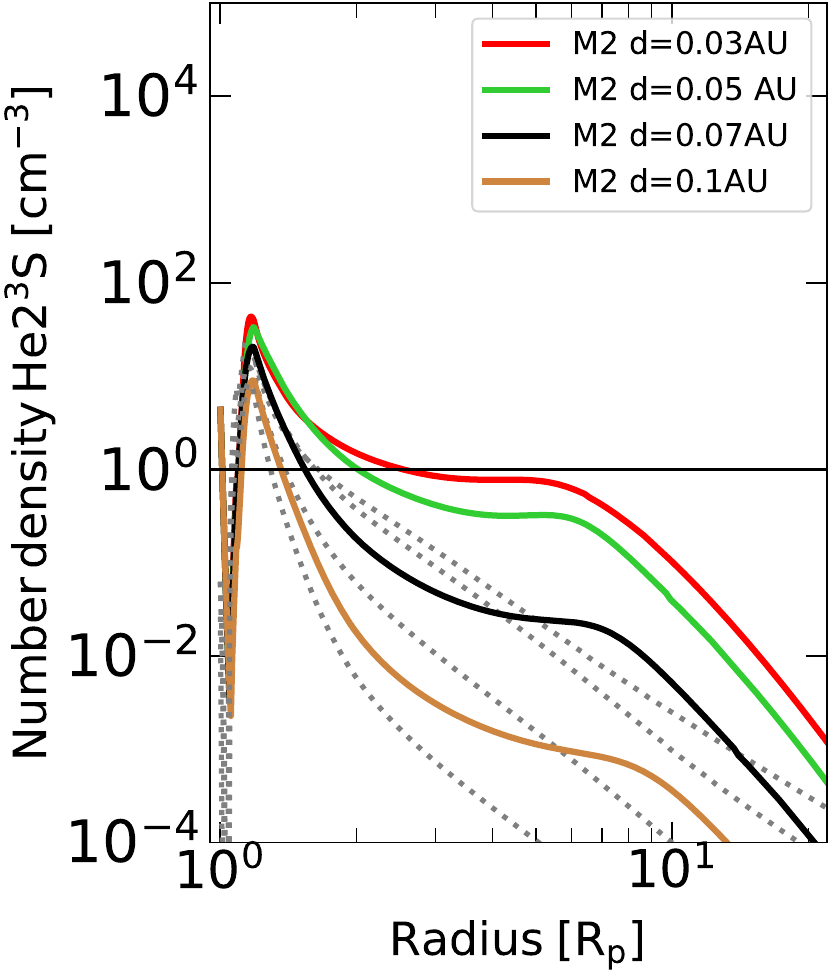}
\includegraphics[width=0.32\textwidth]{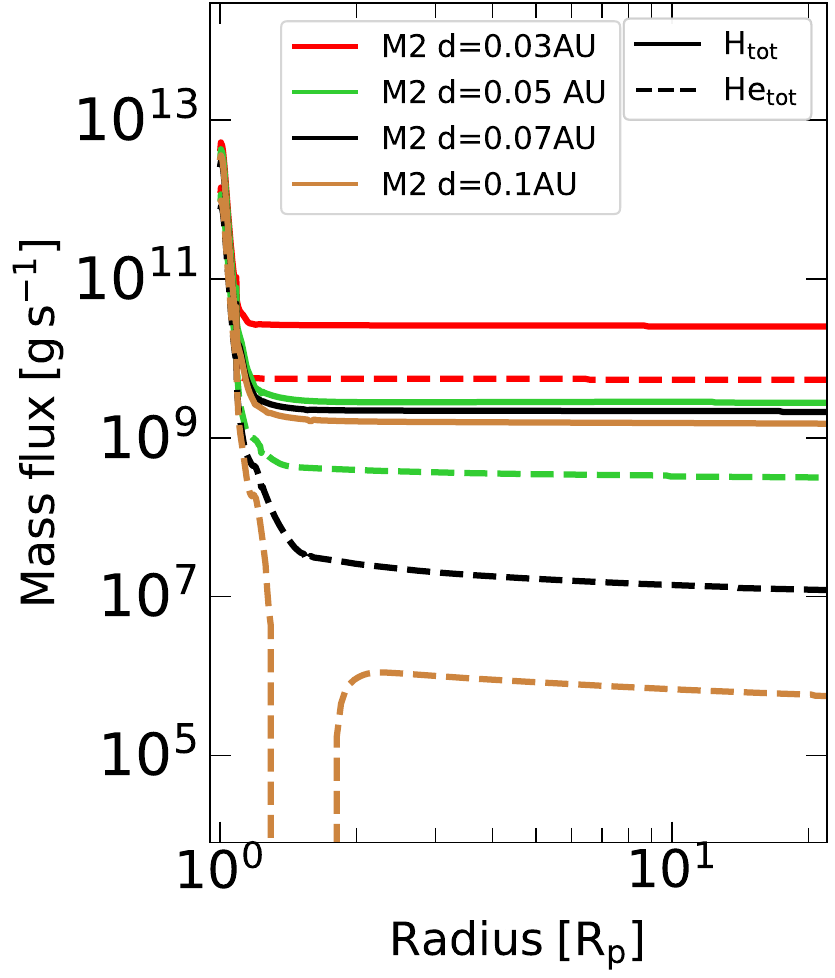}
\includegraphics[width=0.32\textwidth]{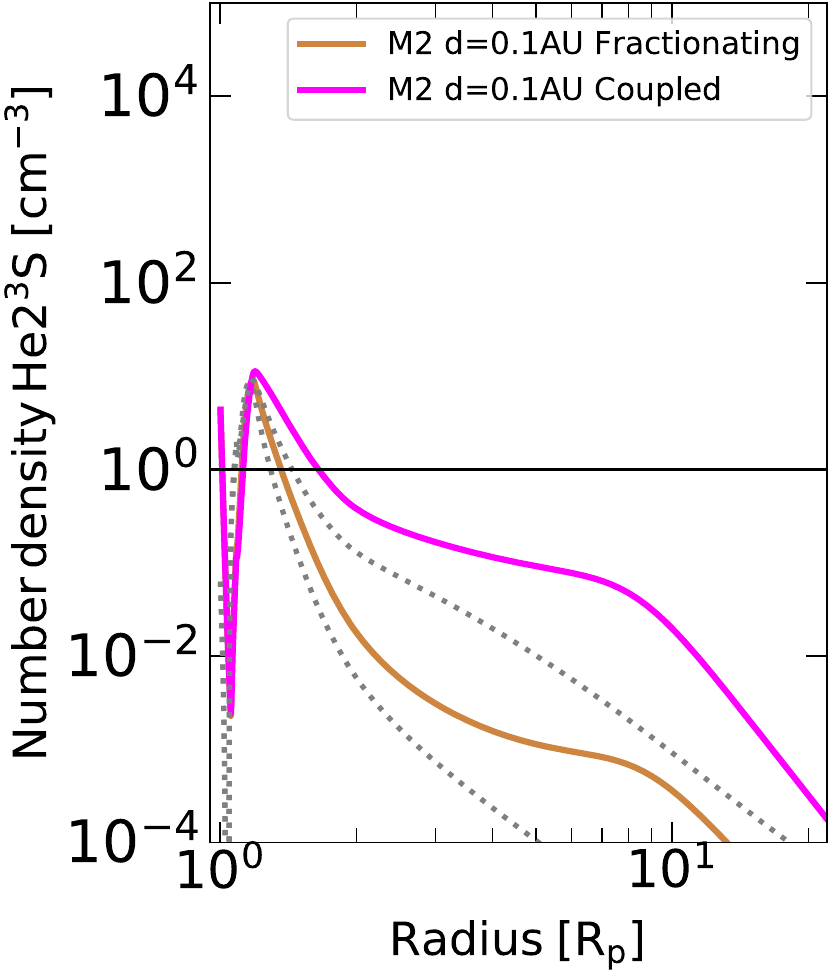}

\includegraphics[width=0.32\textwidth]{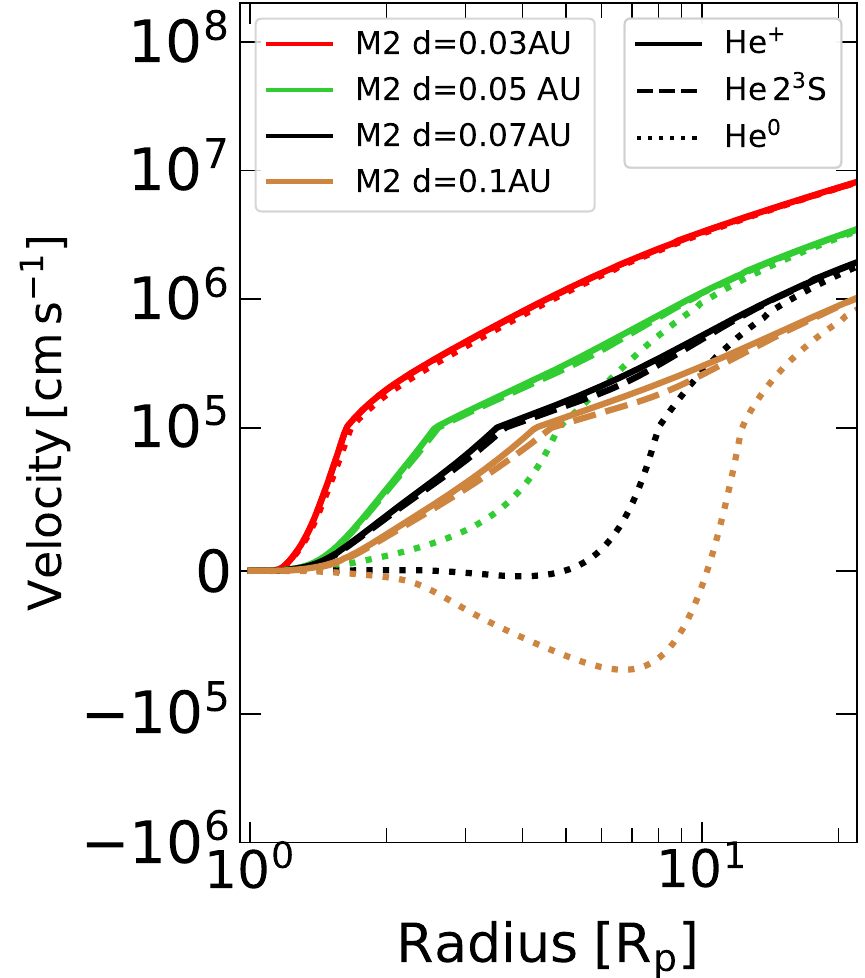}
 \includegraphics[width=0.31\textwidth]{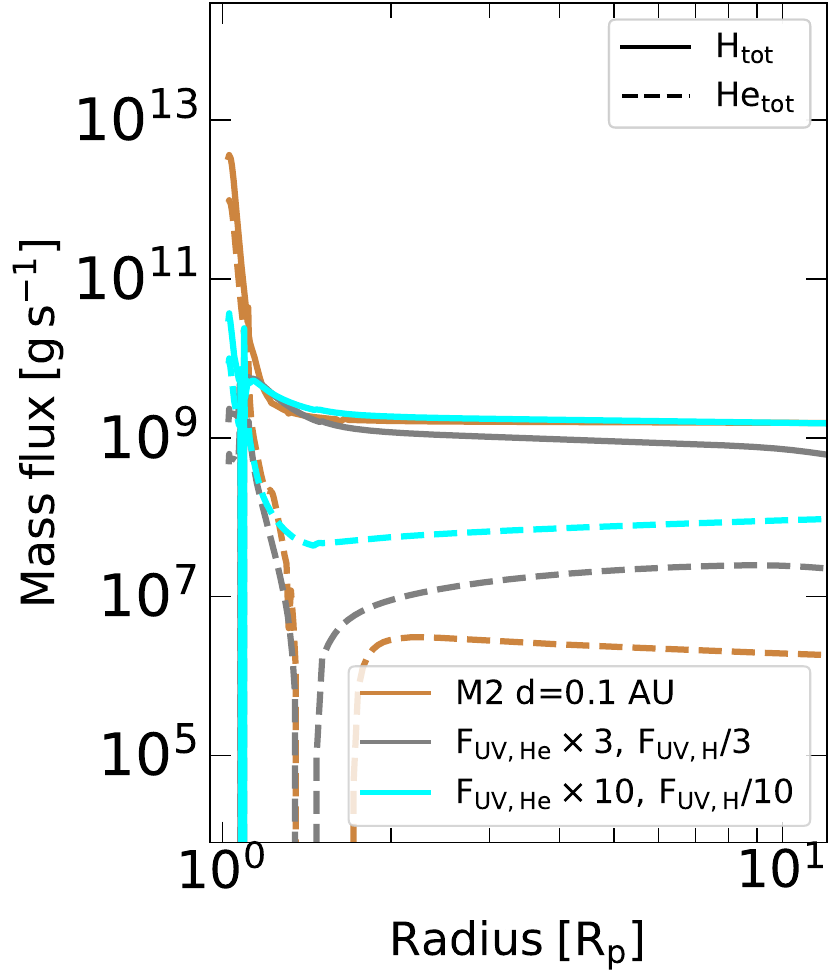}
 \includegraphics[width=0.32\textwidth]{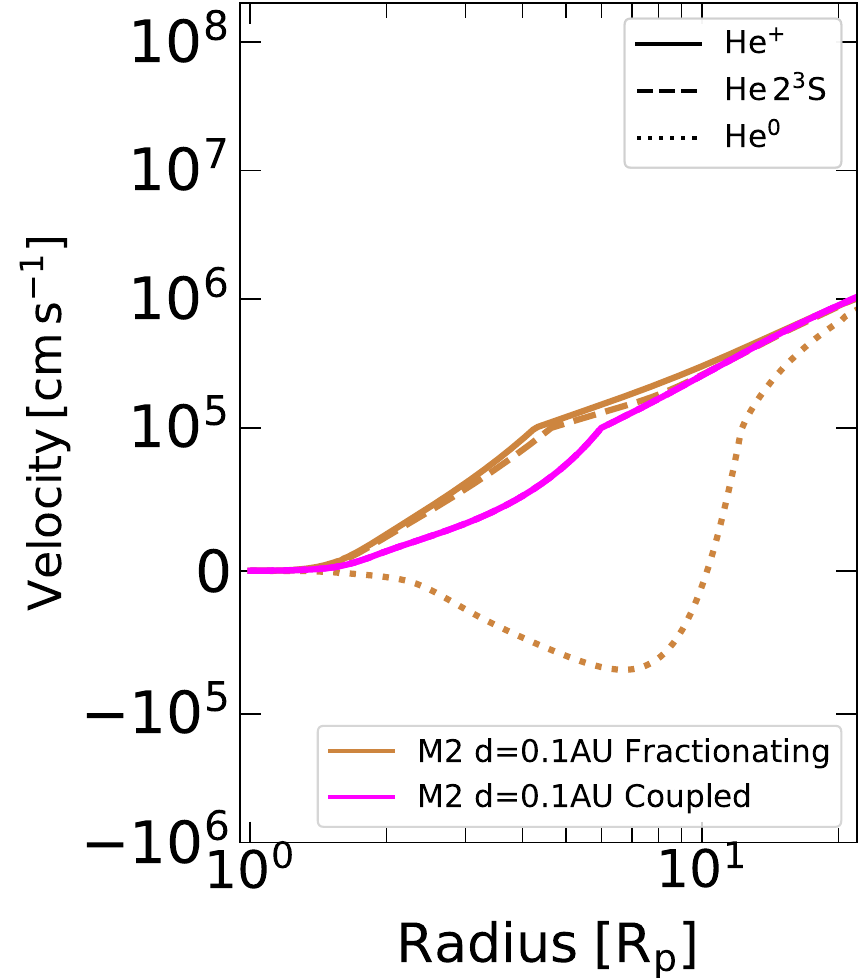}
    \caption{Fractionation and sensitivity of M2 star-hosted outflows on the planet-star separation {and the ionizing flux}: Fractionation sets in at $d=0.05$ AU, which prevents excessive transit depths for larger distances and suppresses most of the number density altogether at $d=0.1$ AU, shown on the left. The top middle panel shows the mass fluxes of neutral and ions together as total flux. This emphasizes that distant planets effectively cease to loose helium.
    Increasing the He-ionizing photon flux, while decreasing the H-ionizing photon flux in the bottom middle panel emphasizes the impact of ionization on fractionation of the total helium mass flux. More ionized flows fractionate less {as can be seen in the central bottom panel, where we increased the helium ionizing flux at $E > 24$eV, but decreased the hydrogen ionizing flux at $24$ eV $> E >13.6$ eV in order to keep the hydrogen mass flux constant, which drives the fractionation}.
    Perfectly coupling all species together, shown on the right, mimics the effect of strong ionization and suppresses fractionation, making the He$2^3$S line detectable.
    The Excess absorption profiles compared to perfectly coupled simulation are shown in Fig. \ref{fig:helium_transits_fractonation_and_widths}. {The grey dotted lines in the top left and top right panels are, as in Fig. \ref{fig:physics_intro_to_triplet}, the isothermal equivalent populations of the metastable state}. }
    \label{fig:velocity_fractionation_M2distances}
\end{figure*}

\begin{figure*}
\includegraphics[width=0.75\columnwidth]{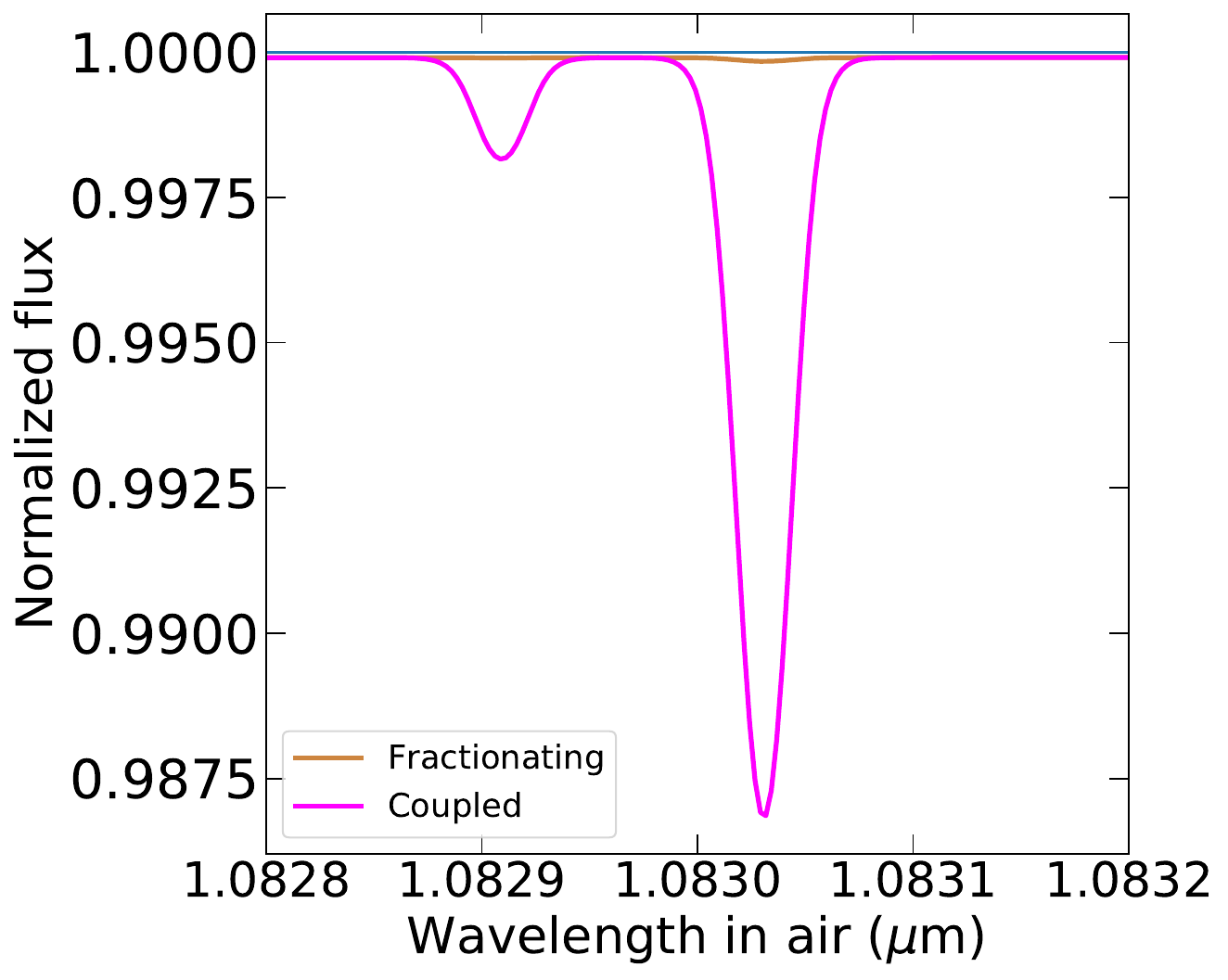}
 \includegraphics[width=0.71\columnwidth]{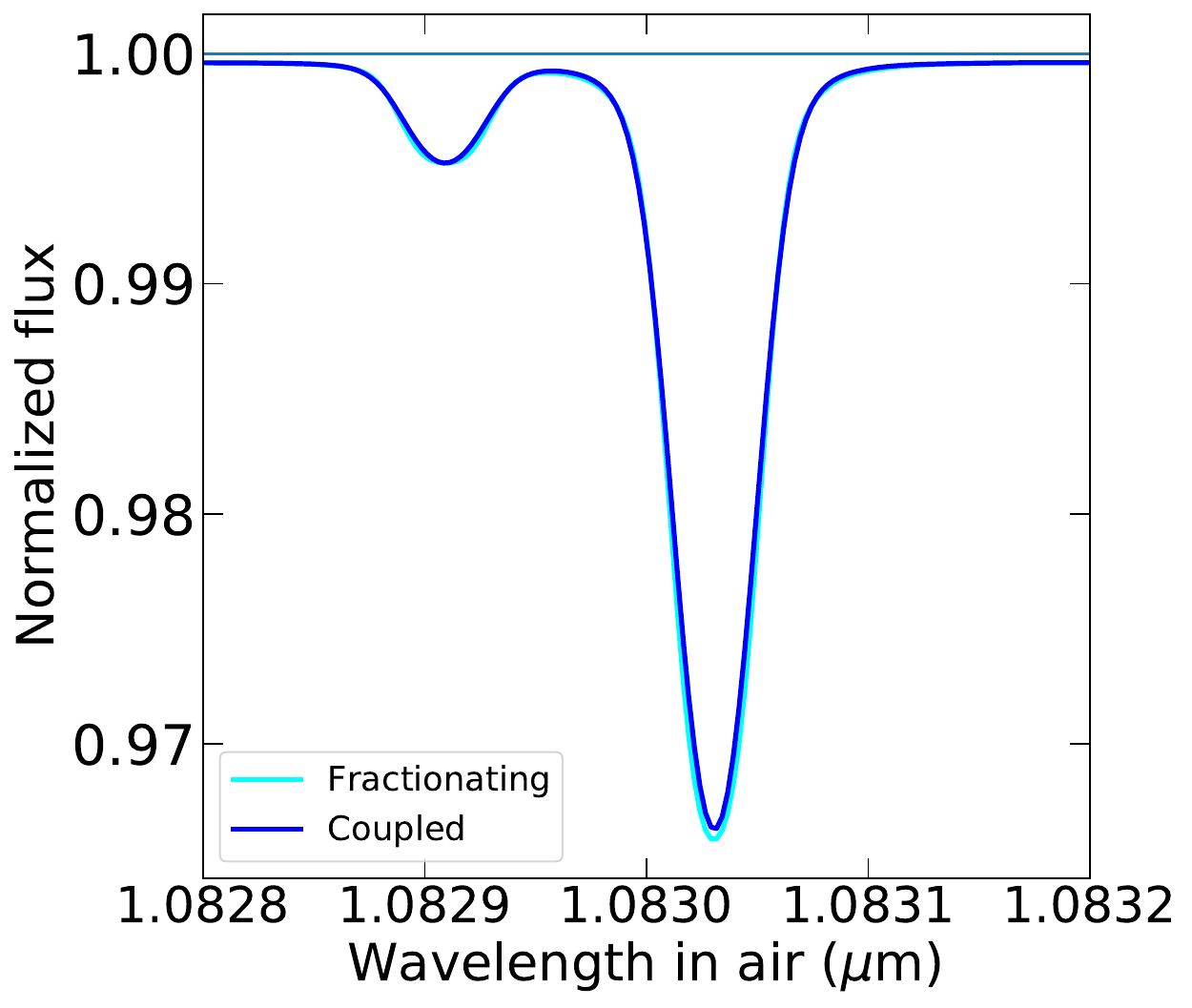}

 \includegraphics[width=0.666\columnwidth]{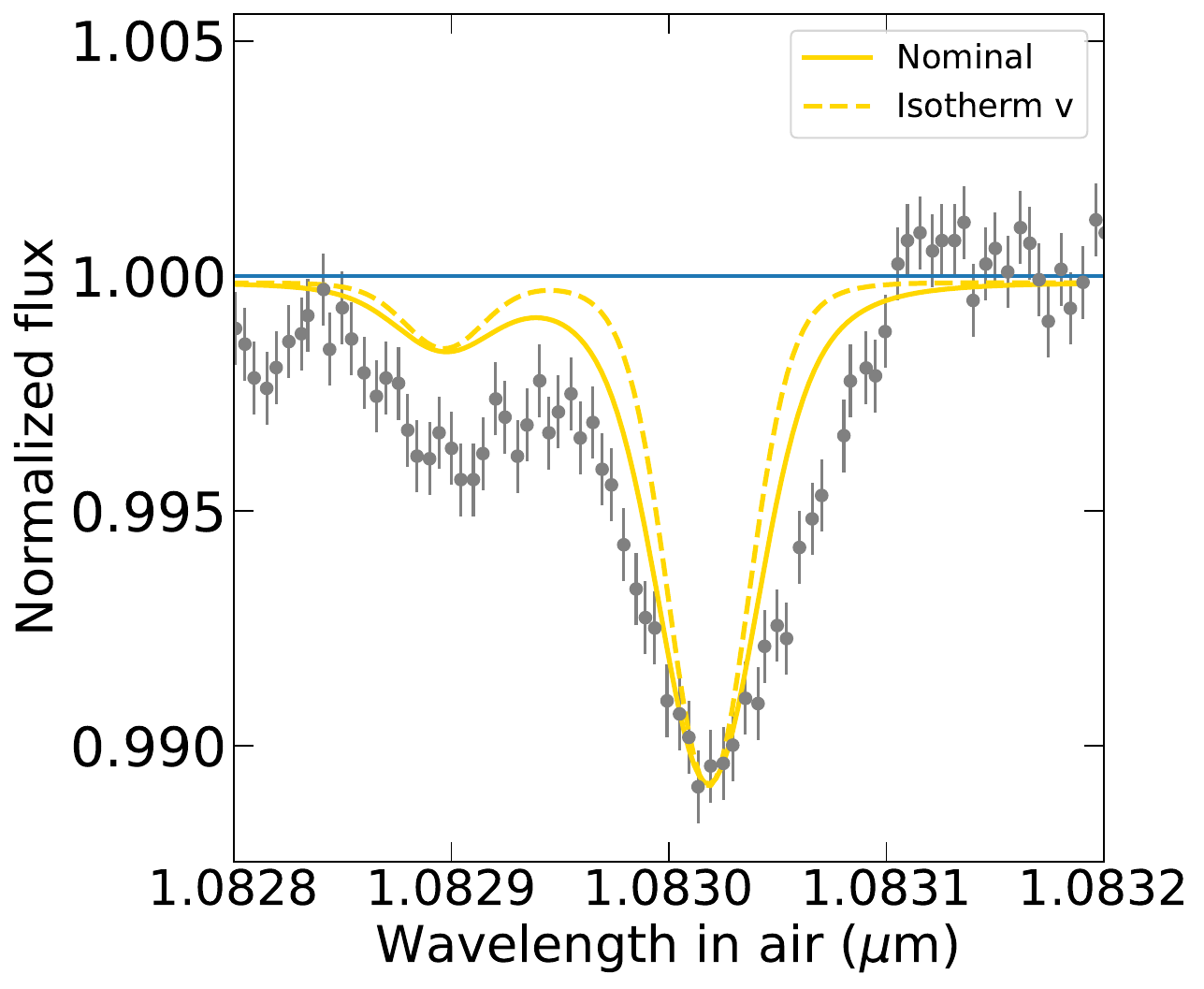}
 \includegraphics[width=0.666\columnwidth]{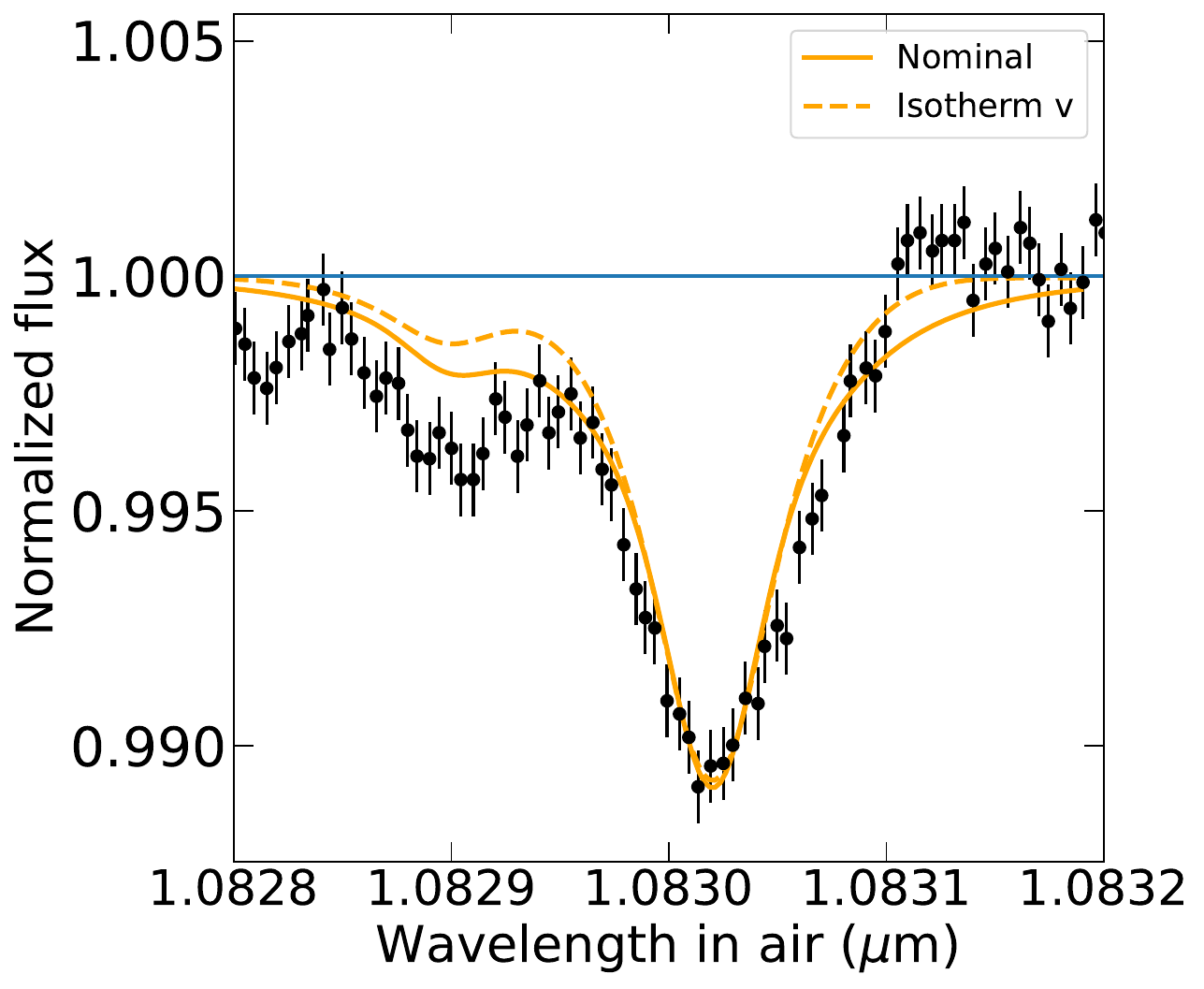}
 \includegraphics[width=0.666\columnwidth]{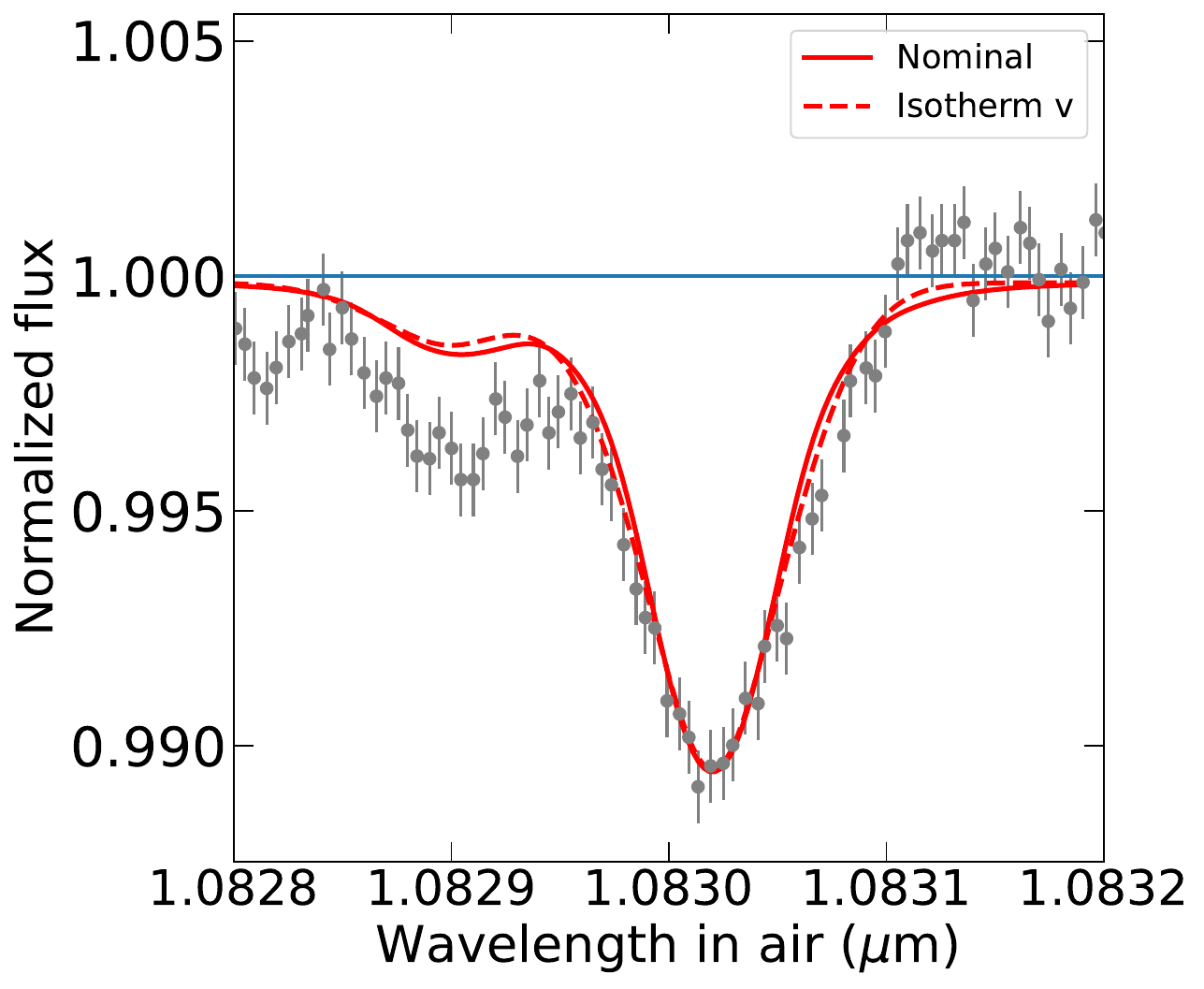}
    \caption{Triplet helium transit profiles demonstrating the role of fractionation (top row), and flow kinematics (bottom). Coupled vs non-coupled simulations (top) for $d=0.1$AU (left) and $d=0.05$AU (right): Fractionation occurs for distant planets around M2 star hosts, whose transits are dominated by natural uncertainty effect broadening, as opposed to velocity-broadening (see bottom plots). This explains why there are only weak fractionation signatures in the transit depths, as long as the ion fraction of Helium is sufficient.
    The bottom plots show the impact of kinematics, for which we used the original simulations from Fig \ref{fig:velocity_fractionation_stars} and compare them to the same number density profile, but with the velocity field replaced by $10^4$K Parker-wind velocity fields. This emphasizes the influence of the velocity field of moving the population bump to higher $\Delta \lambda$). We also overplot the transit data for HD189733b to aid the eye in assessing the simulation's line widths. The meaningful comparison with the actual K2 star is shown as black data points, the  exemplary comparison with other stellar types but the same data is greyed out to avoid confusion. It becomes evident that M type transits are not very velocity dependent, i.e. dominated by natural line width. G and K-types are velocity-dependent due to their higher velocity in the region that produces significant helium absorption.  All simulations had to be slightly rescaled to match the transit depth, scaling factors can be found in the text. }
    \label{fig:helium_transits_fractonation_and_widths}
\end{figure*}

\subsection{Fractionation for M2 hosts}
\label{sec:fractionation_M2}

The sample of helium triplet observations compiled by \citet{dosSantosreview, orellmiquel2024} pointed out a dramatic dearth of detections around M-type hosts. This contrasts with predictions based on the spectra of those stars \citep{Oklopcic2019, biassoni2024}, which would forecast the deepest transits of all stellar types for M stars.


At large star-planet separations, the neutral helium components severely fractionate, even to the degree of developing negative velocities, i.e. flowing back to the planet. 
Incidentally, for those cases Helium is increasingly neutral, and most of it slips through the hydrogen outflow. Because the He$2^3$S population is tied to the number of $\rm He^{+}$-ions, as shown in Eqn. \ref{eq:triplet_rate_solution_simple}, then fractionation becomes visible as clear depression of the He$2^3$S population (Fig. \ref{fig:velocity_fractionation_M2distances}, Left) . 

Moving closer to the star, we find that the onset of He$1^1$S fractionation for the HD189733b analogues is located at a distance of $d\sim$0.05 AU while He$2^3$S remains 'fast' at all distances. 

Some more light can be shone upon this phenomenon by performing some simple numerical experiments. Firstly, we confirm in Fig. \ref{fig:velocity_fractionation_M2distances} (Center bottom) that indeed the ionization state of Helium has a large impact on fractionation. In this experiment, we increase the radiation shortwards of 24$\rm eV$ by a factor of 3 and 10, but decrease the radiation just shortwards of 13.6$\rm eV$ {by the same factor. This choice keeps the outflow rate of hydrogen constant, which drives the main outflow, and together with electrons provides collisions to drag the other species. However, this choice changes the ionization state of helium, which allows us to study the influence of its ionization state on the total helium mass flux. Indeed, as expected, we see that the total hydrogen mass loss rate remains approximately constant, but the total escape rate of helium increases more than linearly with the flux in the spectral bands capable of ionizing it}.

Secondly, we consider the impact of coupled velocities, an approach which is also used in the majority of previous modelling efforts {and analogous to the previous experiment as it effectively sets all collision rates to infinity}. We run an additional simulation which suppress fractionation by forcing all species to attain the common center-of-mass velocity at every timestep. We name this simulation 'coupled' and our nominal simulation approach is named 'fractionating'. The results for the number density profiles and the velocity structures can be seen in Fig. \ref{fig:velocity_fractionation_M2distances} (Right).
It becomes clear that at $d=0.05$AU, there is only a negligible difference between the two simulation approaches. This stems from the fact that ions are always well-coupled to the outflow, and at this distance most of the Helium is ionized. We do see a stark discrepancy between the approaches however for the $d=0.1$AU simulations, because only the coupled simulation drags neutral Helium efficiently into the outflow. { At $d=0.1$AU both simulations possess high neutral fractions at the base of the outflow, leading to fractionation in the uncoupled simulation. Ultimately this leads to a deficit of $\rm He^{+}$ and hence $\rm He2^3S$ in the outflow, explaining the large deficit in the metastable population compared to the coupled simulation}. We further note that the Hydrogen outflow driving the Helium mass-loss decreases accordingly with distance, contributing to the fractionation of neutral He.

We summarize the physics found in this section with a sketch of their observational consequences. Because the He$2^3$S population remains fast due to inheriting the $\rm He^{+}$-velocities, the expected line broadening is that given by the overall outflow velocity, even for fractionating flows. However the line depth, given by the number density of $\rm He^{+}$ can vary significantly, depending on the ionization state of the outflow and the hydrogen mass-loss rate driving said flow.




 

\begin{figure*}
 \includegraphics[width=1.2\columnwidth]{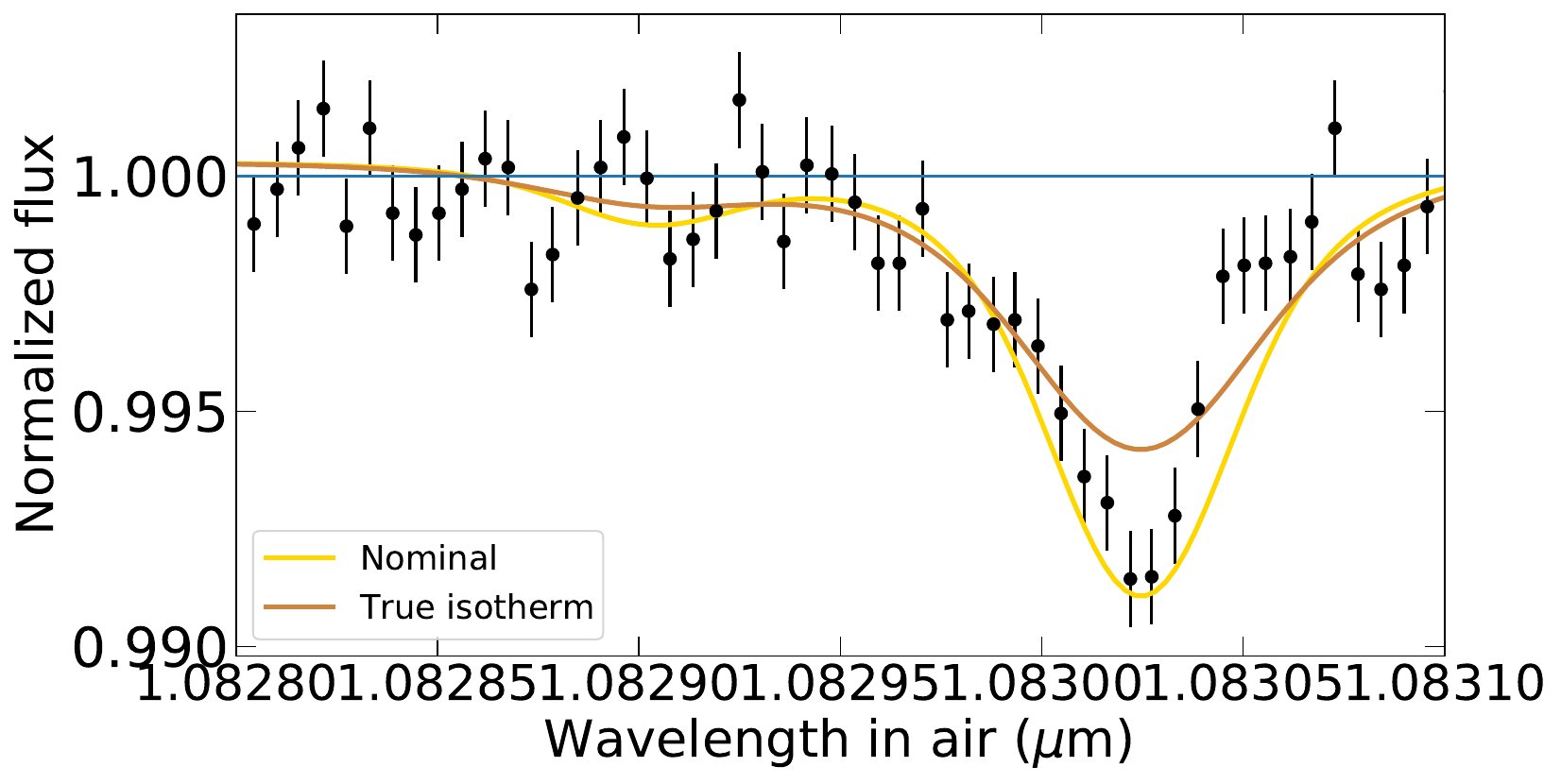}
  \includegraphics[width=0.68\columnwidth]{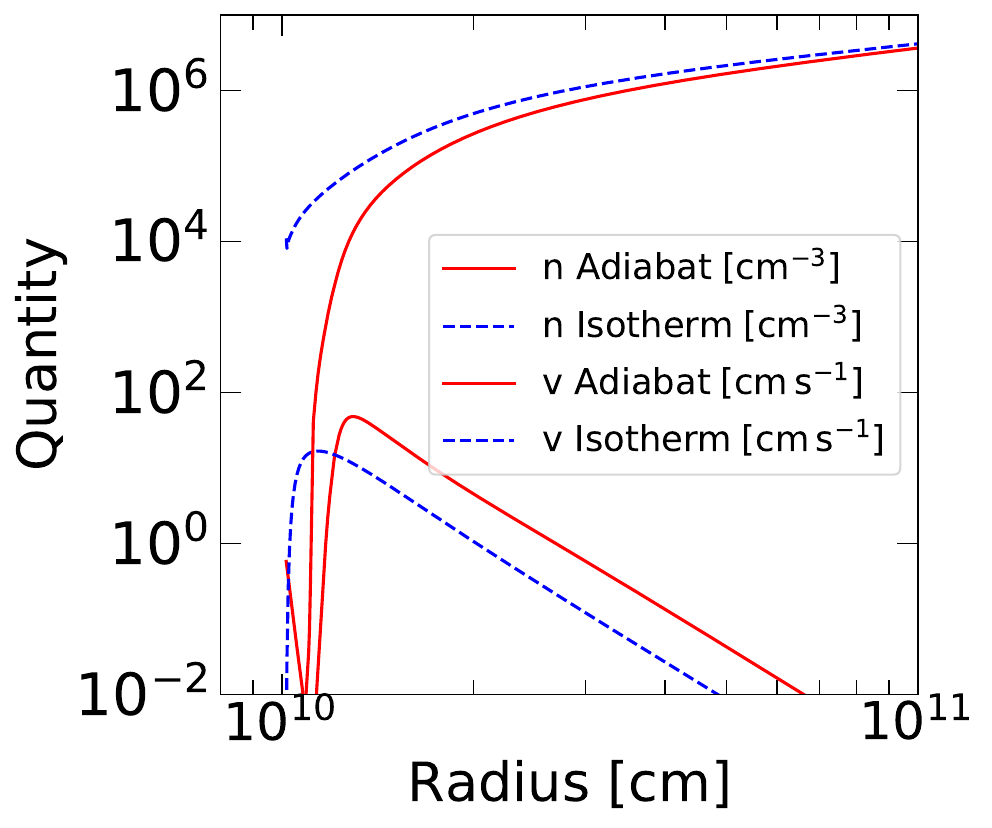}
 
    \includegraphics[width=1.3\columnwidth]{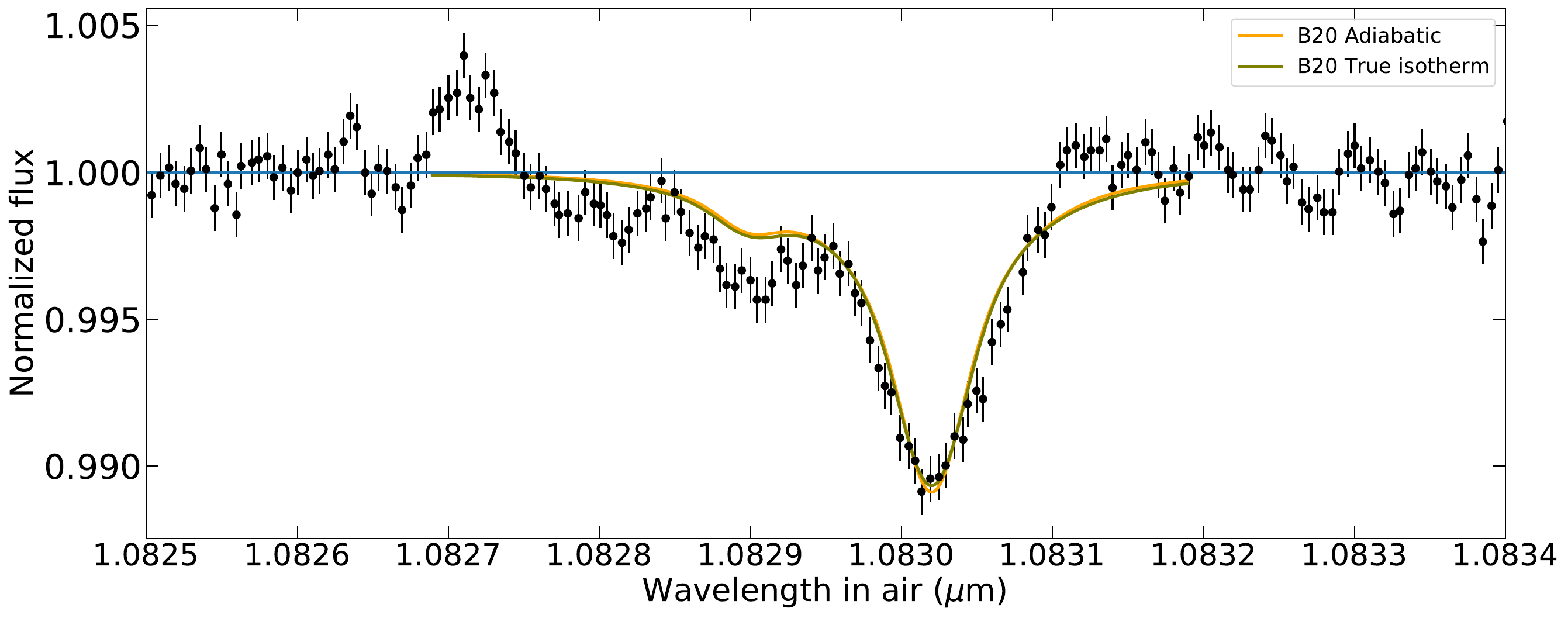}
  \includegraphics[width=0.68\columnwidth]{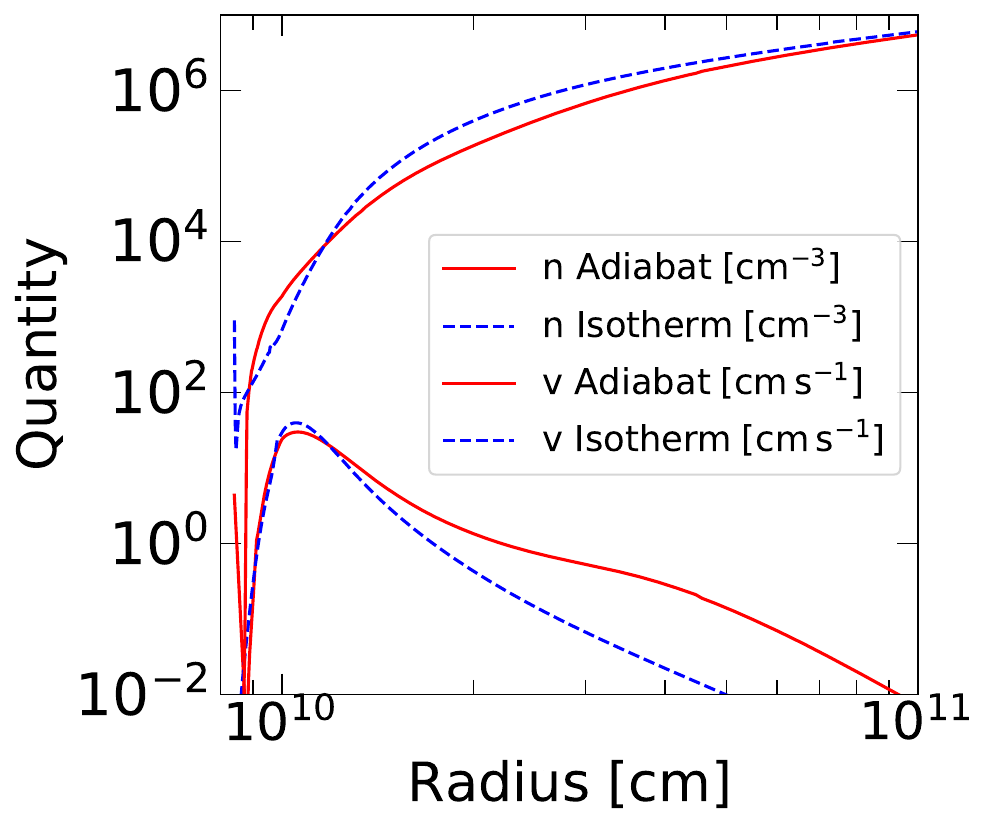}

\includegraphics[width=1.3\columnwidth]{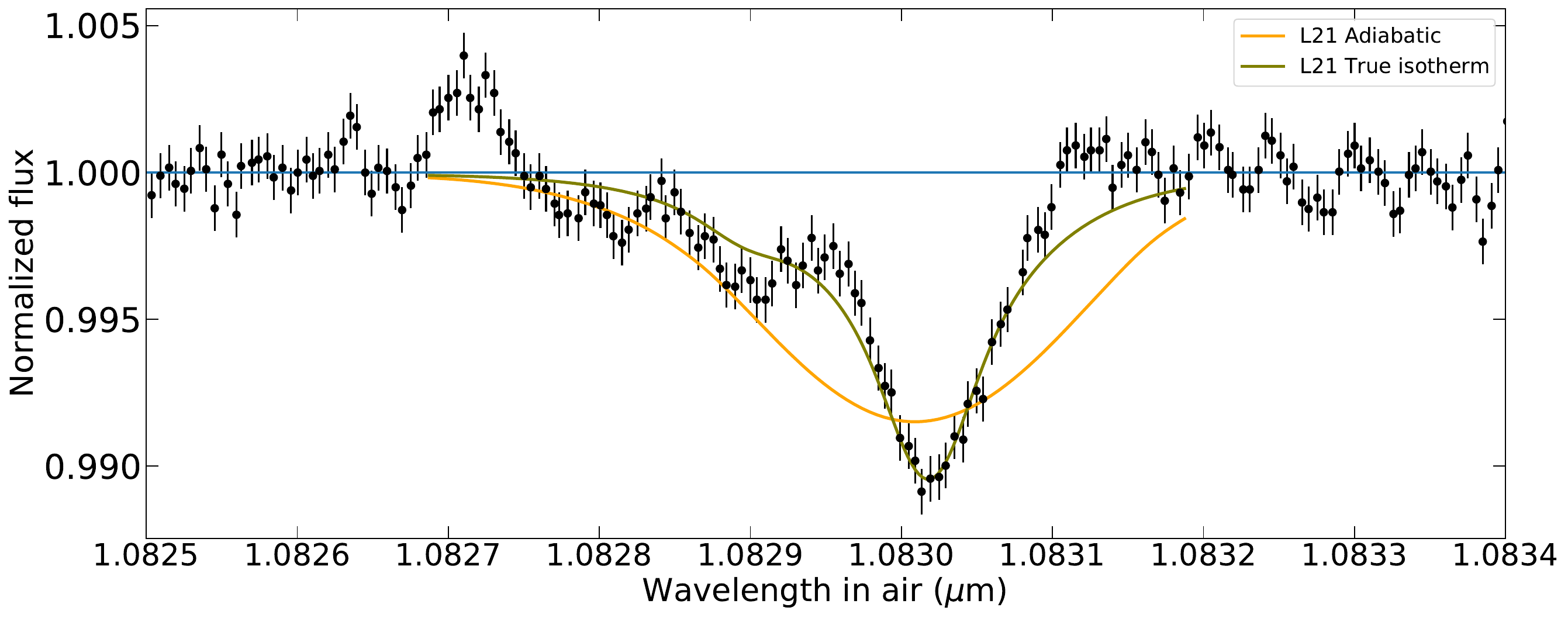}
  \includegraphics[width=0.68\columnwidth]{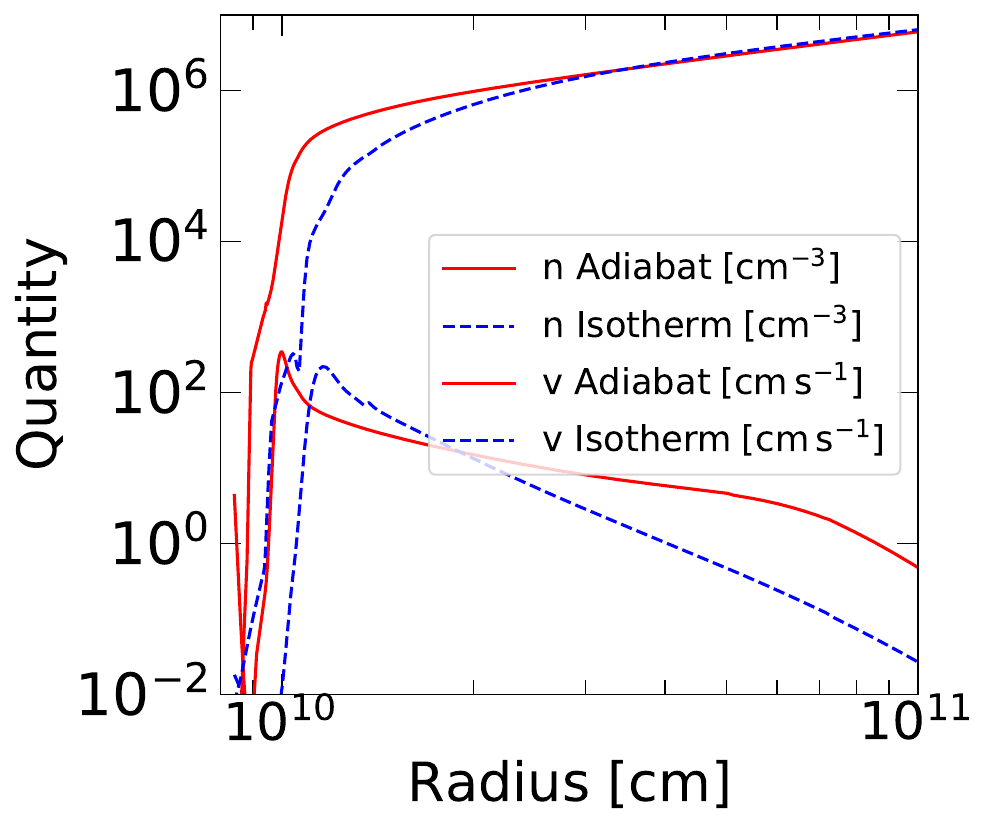}

  \includegraphics[width=1.3\columnwidth]{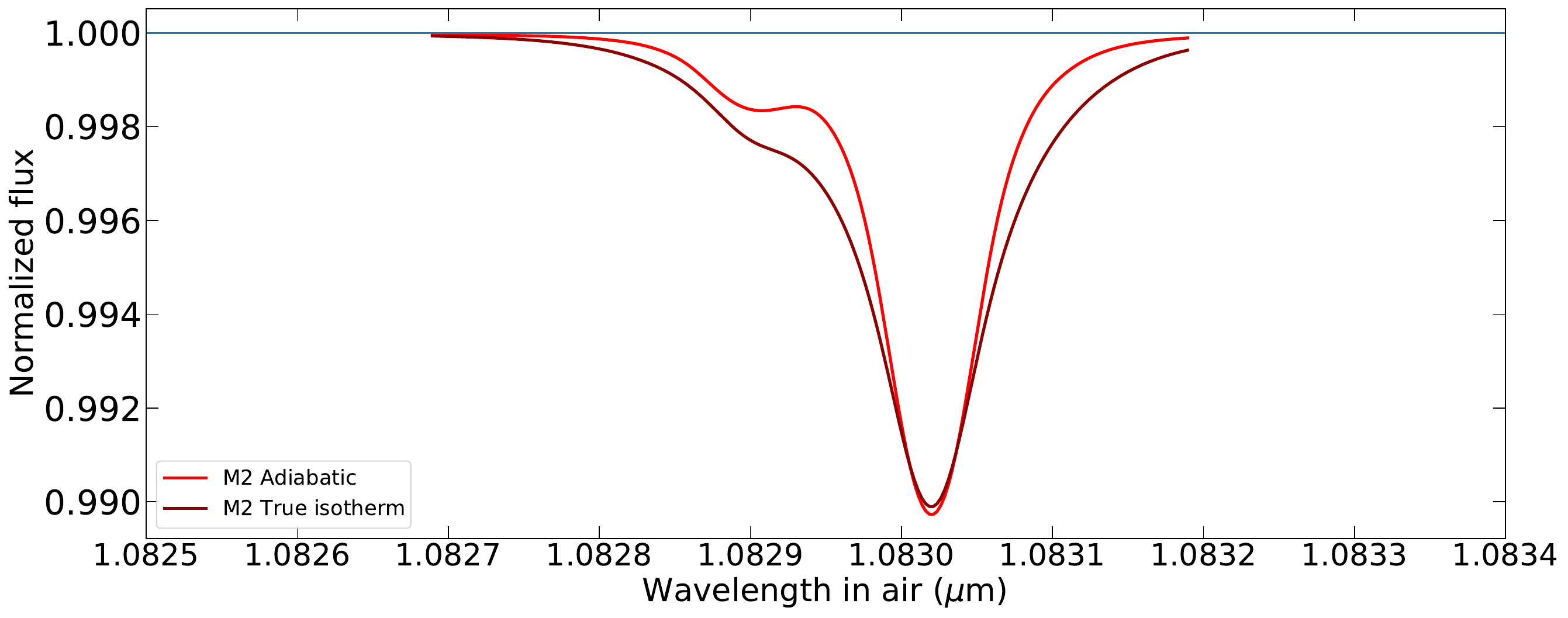}
  \includegraphics[width=0.68\columnwidth]{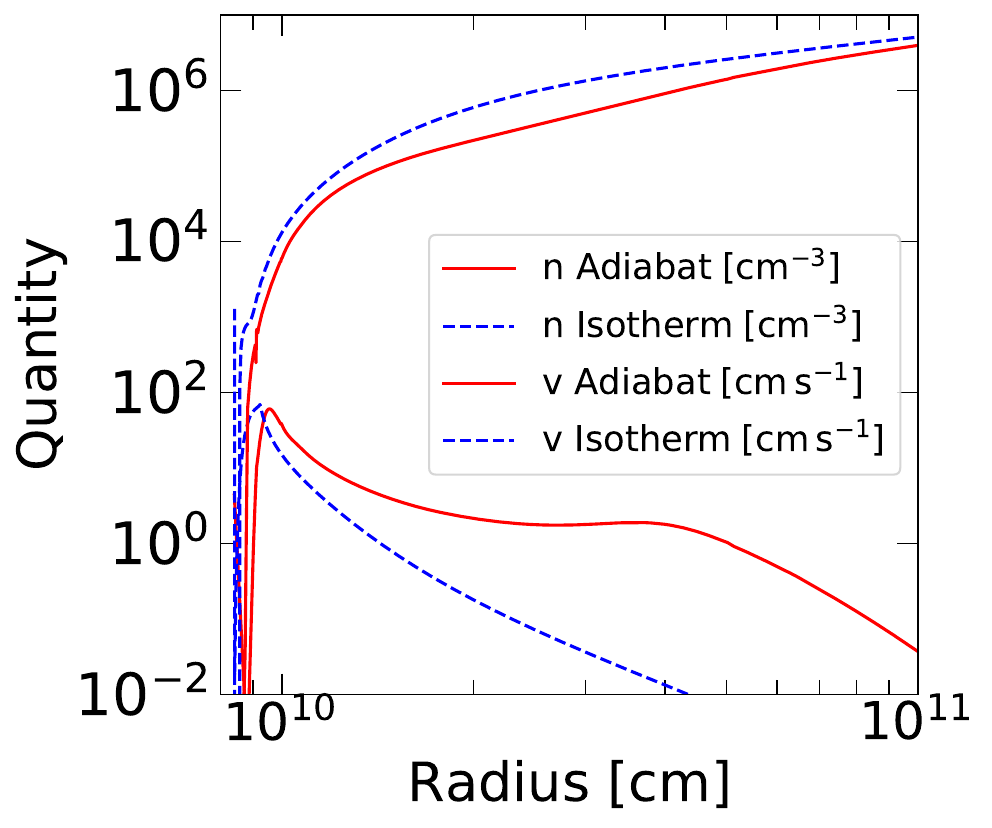}
 \caption{Helium transits profiles (left) and the outflow properties (right) for our self-consistent simulations and equivalent isothermal Parker wind solutions.  \textbf{Top:} The models for HD209458b. Due to the suppression of the triplet population bump, it is difficult to assign line broadening to any individual physical process. \textbf{Second row:} HD189733b irradiated by the \citep{bourrier2020} spectrum for HD189733. Similar velocity solutions conspire to give similar line broadening, but the absolute amount of He$2^3$S can be used to distinguish the models. \textbf{Third row:} HD189733b, irradiated by the \citep{lampon2021} spectrum for HD189733. The higher mass-loss rate and faster outflow demonstrates the sensitivity of the triplet line around this planet to the velocity structure on which the population bump is attached to. \textbf{Bottom:} Our M2 simulations, without data, rescaled to the same transit depth. The isothermal simulation is much broader due to strong velocity broadening, despite the absence of the population bump. }
    \label{fig:helium_transits_data_realcomparisons}
\end{figure*}

\section{Transit predictions}
\label{sec:linemodelling}Following previous work, we assume our outflows generate a spherically symmetric density profile, which we simulate out to $30r_{p}$, which is our outer simulation boundary, in order to compute the transit signatures resulting from our runs. We then take our temperature profiles and triplet density profiles and make use of the transmission spectra simulation tool {\sc p-winds} \citep{p-winds}\footnote{A recent update of {\sc p-winds} allows transit simulation through non-isothermal density profiles}. While {\sc p-winds} is capable of generating isothermal velocity profiles based on the Parker wind solution (similar to the \citealt{Oklopcic2018} model), we only use it to calculate the transmission spectra from our self-consistent {\sc aiolos} models. 
 

\subsection{Impact of velocity profiles: Perfect coupling and fractionation}


In this section we use our simulations to estimate the observational effects of fractionating atmospheric outflows. Given the deep gravitational wells of giant planets, this is a maximum estimate. A focus on Sub-Neptunes and the radius valley \citep{Fulton2017} is relegated to future work. 

As discussed in the previous sections, even though ground-state neutral helium can decouple from the outflow, we showed in Fig. \ref{fig:velocity_fractionation_M2distances} that the velocities of the ionized bulk escaping gas can be imprinted onto the triplet state. This means that the triplet state is significantly less prone to fractionation; thus, previous work employing perfect coupling between all species' velocity components is a reasonable assumption 
\citep[e.g.][]{biassoni2024}.

Discrepancies between predictions of a perfectly coupled simulation and a fractionating simulation are mainly expected once the neutral Helium fractionates so strongly that it develops negative velocities (i.e. it becomes inflowing), and the tidal field of the star is insufficient to accelerate it to the same velocity as the bulk flow. Essentially, the negative velocities can then close a cycle where ionized helium is outflowing; it then recombines, falling back towards the planet before the same atom can be ionized again and it re-enters the outflow \citep[as also found in][]{xing2023}. 
This situation would be expected to be impactful for planets at large semimajor-axis distances, which is why we investigate the transit prediction for the perfectly coupled and a fractionating simulation of the M2-hosted giant at $d=0.05$ AU and $d=0.1$ AU, which we show in Fig. \ref{fig:helium_transits_fractonation_and_widths}. 

Our findings for the predicted transit signatures follow the expectations laid out in Section \ref{sec:fractionation_M2}. In comparing fractionating and coupled simulations, the $d=0.05$ AU simulations show no difference, due to similar number density and velocity profiles. The transit depth of $\sim 3\%$ is furthermore similar to the findings in \citep{biassoni2024}, where those authors find $\sim 4\%$, albeit for a different stellar spectrum.
However, the case of the more distant $d=0.1$AU simulations, which would result in a $\sim 1\%$ transit depth for the coupled simulation, shows a negligible signal in the fractionating counterpart, as expected from the extremely low number density in this simulation.
The sudden onset of this process might be surprising, but as previously mentioned, ionization and the drop in Hydrogen mass-loss rates conspire to decrease the amount of He$2^3$S available. In order to suppress the He$2^3$S transit signal, the flow has to be fractionating and neutral.


Thus we conclude that Helium fractionation is a possible candidate to explain some triplet nondetections around giant planets.
However most of the non-detections of escaping helium around M-dwarfs stem from planets smaller than $5\, R_{\oplus}$ \citep{orellmiquel2024}, i.e. Sub-Neptunes.
It remains to be seen if the outflows of Sub-Neptunes like GJ 436b \citep{Ehrenreich2015}, which are easily detected in the neutral Hydrogen Ly-$\alpha$ line, follow the same physics outlined here, or whether there are other factors at play, such as radiation pressure \citep{rumenskikh2023}.


\subsection{Impact of velocity profiles: Nonisothermal vs. isothermal velocities - "Sensitivity of the bump contribution to velocity line broadening"}

We now focus on the transit signatures of non-fractionating giant planets at $d=0.03$ AU, emphasizing the impact of different treatment of thermodynamics.

Hydrodynamic outflows are expected to cool adiabatically, with adiabatic cooling being particularly important for giant planet outflows \citep{MurrayClay2009, salz2016, Caldiroli2021}.  In these outflow structures, a helium triplet population maximum naturally occurs \citep{yan2022}. Thus, it is important to characterize the observational signature of these flows over those of isothermal flows, which are often used to compare to observations \citep{lampon2021}. There are two important aspects to this comparison, which contribute differently to the broadening of the Helium triplet line: firstly, will isothermal flows not featuring a triplet population maximum have excessive thermal broadening at large radii compared to their adiabatic counterparts?
Secondly, flows featuring the adiabatic population bump might not show up as anomalously broadened - as the bump needs to be located at large velocities to be distinct from a simple deep transit at the line center. We first investigate the latter effect in this subsection by comparing two possible realistic velocity profiles - self-consistently calculated and isothermal - both transporting the same population bump outwards. This allows us to characterize which planets are sensitive in their transit signals towards the velocity profile. The effects of the temperature profile on broadening itself is discussed in the following section. 



Using a fixed triplet number density profile, we use the simulations at $d=0.03$ AU, shown previously in Fig. \ref{fig:effect_of_stars_on_triplet}, and vary the velocities between the isothermal and self-consistent profiles. 
Those simulations do not fractionate, i.e. $v_{\rm He2^3S}=v_{\rm bulk}$ and hence the triplet velocity profile can be meaningfully compared to analytic single-fluid solutions, such as the Parker wind profile \citep{cranmer2004}. The latter is another 'maximal' velocity assumption, as the tide-less \cite{cranmer2004} velocity profile accelerates much more steeply than the tides-including isothermal velocity profile as solved by \citep{vissapragada2022}. The chosen isothermal temperature to generate the isothermal velocity profiles is $T=10^4$K, which is a typical choice in the literature \citep{lampon2020, lampon2021, Zhang2023}. Furthermore, since above 10$^4$~K, Lyman-$\alpha$ and recombination cooling are extremely strong, any outflow is thermostated back to 10$^4$~K \citep[e.g.][]{MurrayClay2009,owen2016}. Hence, we can consider this temperature choice to represent the maximal thermal broadening width an outflow could achieve.

We show this comparison in Fig. \ref{fig:helium_transits_fractonation_and_widths} (Bottom). Additionally, we overplotted the transit data for HD189733b from \cite{salz2018} with all three \ms{simulated stellar host} cases to provide a reference (we are not trying to ``fit'' HD189733b here). As those simulations, when run through \textsc{p-winds}, give a different transit depth from the measured data, we rescale our helium densities by factors of $(1.4,\, 1.34,\,0.33)$ for the $\rm (G2,\,K2,\,M2)$-star. The spectrum used for the K2 star is the one by \cite{bourrier2020}, we will discuss the observational consequences of using \cite{lampon2021} below.

The correction factor for the K2 star of $1.34$ could indicate that we might be overestimating the fractionation in K2 hosts. { This suspicion arises because the current fractionation factor in this simulation is 0.20. In a perfectly coupled, non-fractionating flow with the same helium content, this would be consistent with our set primordial helium content as it is $0.2 \times 1.34 \approx 27 \%$.}

\ms{Because those three stellar cases naturally produce a span of different mass-loss rates and velocity profiles, we now can use this data to investigate the sensitivity of to the helium population bump to the underlying velocities:}

\begin{itemize}
    \item \textbf{G2 stars}: There is a noticeable difference between using the self-consistent and isothermal velocity profiles. 
    {However, for this stellar type we caution that due to the absence of a population bump in both isothermal and adiabatic simulations, the broadening is dominated by Doppler broadening. While a comparison between G2 isothermal and G2 self-consistent might find a broadening difference, this cannot be due to the population bump, as it is absent in both.}
    {This velocity broadening effect can also be achieved by inappropriate inner boundary conditions for the density and pressure.} Setting a $P\sim$nbar boundary condition at the transit radius, which should  represent the $P\sim$mbar radius, is not appropriate. {Seemingly small errors in this boundary condition can lead to large differences in the adiabatic velocity structure as function of radius.}
    \item \textbf{K2 stars}: 
    Those outflows feature similarly large velocities at large radii as $G2$-hosted outflows, but the population bump is present. This broadens the transit profile from the self-consistent simulation over the isothermal case in the line wings. This is a key signpost that might be used to hunt for signatures of adiabatic cooling.
    Conversely, we remark that a well-fitting isothermal model is therefore not an indicator that the atmosphere is isothermal.
    \item \textbf{M2 stars}: Here, the transit signal probes the same velocity region for both treatments of thermodynamics, making the line insensitive to velocity. This is unlikely however to be a general result, a case-by-case assesment of sensitivity for other planets is likely to be necessary.
\end{itemize}

 We conclude that K stars, with HD189733b as a prime example, are possible testbeds for detecting adiabatic cooling, if their line wings can be resolved at a high signal-to-noise ratio. K stars are theoretically favoured targets for helium signals \citep{Oklopcic2019}, a conclusion empirically verified \citep{dosSantosreview,orellmiquel2024}. Thus, studies of escaping helium atmospheres of planets hosted by K stars could be crucial quantitative test-beds of photoevaporative driven mass-loss, particularly if performed at high enough spectral resolution to resolve the excess broadening in the line-winds. The corollary of this conclusion is that isothermal fits to the observational data will tend to overestimate the outflow temperature when compared to self-consistent calculations. Such a trend between self-consistently determined and fit temperatures is already present in observations analyzed by \citet{Linssen2024}. For TOI-2134 b they suggested additional broadening mechanisms to explain the fit temperature exceeded that predicted from self-consistent calculations by a factor of $\sim 2$.

\subsection{Detectability of non-isothermal over isothermal structures}
We now compare fully isothermal simulations with self-consistent adiabatic simulations for our best-guess HD209458b and HD189733b simulations. The data for HD209458b is taken from \citep{alonsofloriano2019}. For the isothermal simulations, we fix the mass-loss rates in hydrogen to those found from our adiabatic runs, and fix the isothermal temperatures to $10^4$K. The isothermal simulations are also performed with {\sc aiolos} but reproduce the Parker wind solutions \citep{schulik2023} in tidal fields, as given in \citep{vissapragada2022}.

We note that the isothermal model is too fast and too broad to fit the data, see Fig. \ref{fig:helium_transits_data_realcomparisons} (Top), indicative of temperatures colder than $10^4$K, similarly to previously employed isothermal solutions \citep{lampon2020}.
However, similar to the HD189733b analogue hosted by a G2 star, shown previously in Fig. \ref{fig:velocity_fractionation_stars}, the line width for the real HD209458b, which orbits a G-type star, is mainly sensitive to the velocity structure of the outflow and the amount of helium in the outflow, as the temperature-dependent population bump is suppressed. The transit simulation results require Helium population scaling factors of $\times 4$ for the adiabatic and $\times 16$ for the isothermal simulation. Those trends go into the opposite direction of previous claims of reduced Helium in outflows \citep[e.g.][]{lampon2021}, hence we caution to overinterpret the results from isothermal simulations. \citep{xing2023} also favour reduced Helium abundances in the outflows of HD209458b, albeit at a higher mass-loss rate and fixed heating efficiency. Thus, despite HD209458b being a well-studied example of an exoplanet undergoing atmospheric escape, the outflows of G-type hosted planets are not the best test bed for detailed simulations of wind launching, as the helium triplet signal is not sensitive to the flow thermodynamics. {Finally, we see in many isothermal model comparisons that the density profiles need to be decreased below the adiabatic absorption radius to match the isothermal mass-loss rates to the adiabatic ones. This will ultimately compensate for the onset of helium fractionation in isothermal models.}

\begin{figure*}
\includegraphics[width=1.0\linewidth]{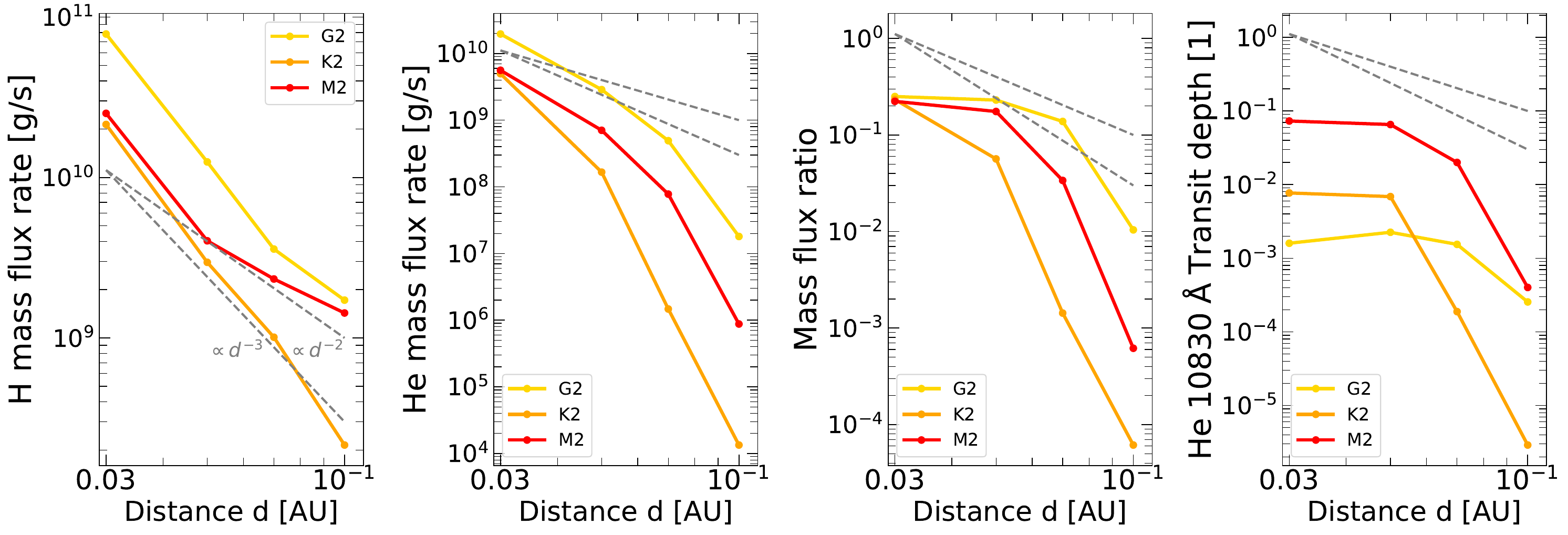}

\includegraphics[width=0.28\linewidth]{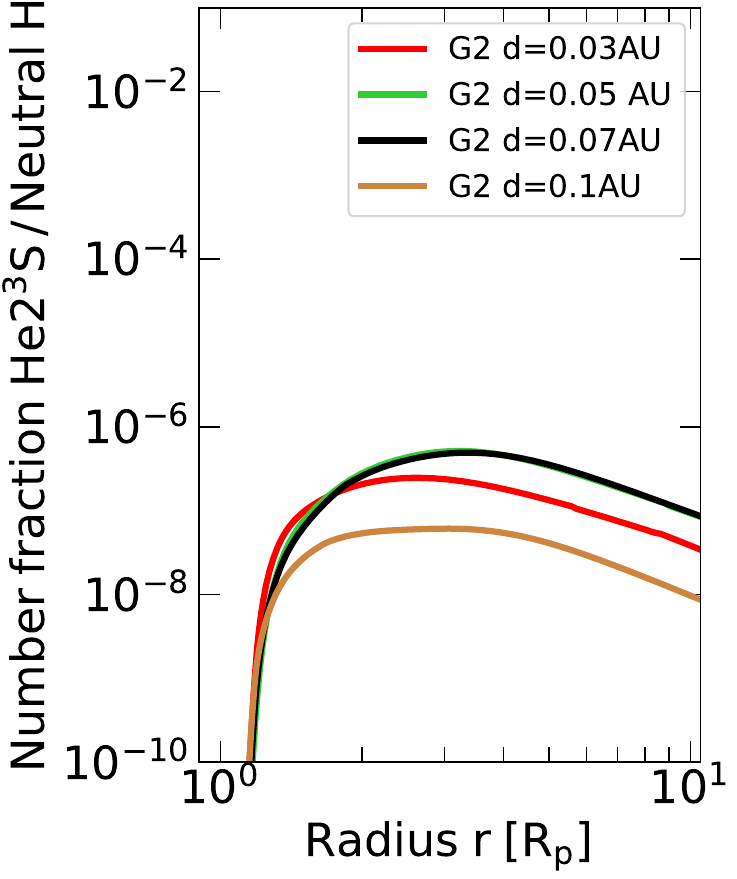}
\includegraphics[width=0.21\linewidth]{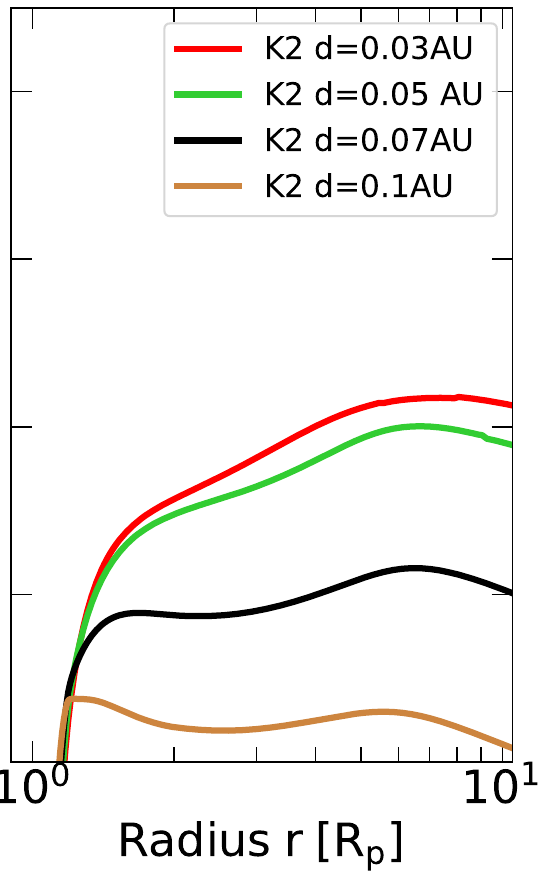}
\includegraphics[width=0.21\linewidth]{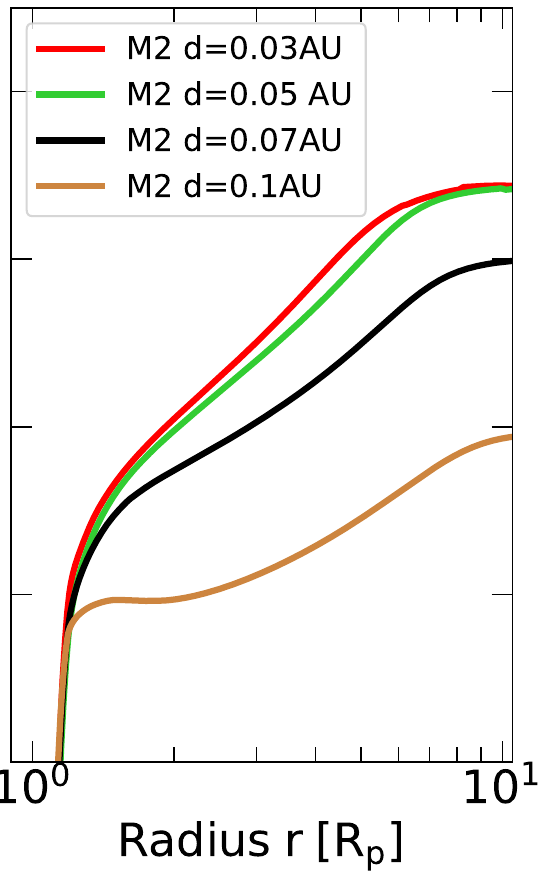}
 \caption{Semimajor axis distance variation for all three spectral types irradiating HD189733b analogues. \textbf{Top}: Interesting quantities as function of planet-star distance. Mass-loss rates in Hydrogen and Helium, as well as their ratios indicating fractionation, and the predicted transit depths, as function of planetary semimajor-axis distance. The dashed grey lines indicate scaling as $\propto d^{-2}$ as would be expected from purely radiation-driven outflows and scaling as $\propto d^{-3}$ as would be expected from tidally driven outflows. Stronger mass-flux scaling indicates the contribution of tidal fields, which decrease in importance for lower mass-stars. Mass flux ratios differing from the primordial $27\% $ indicate separation of helium from the main flow via fractionation. The transit depths roughly follow the mass flux ratios, modified by the overall abundance of $\rm He 2^3S$ due to spectral variations. The transit depth normalization is relative to the out-of-transit light curve, i.e. the close M2 giant should show a $\sim 7-8\%$ transit. \textbf{Bottom}: The ratio of triplet helium atoms to neutral hydrogen as function of planetocentric radius, in the outflows of the planets in the top row. These demonstrate the effect of fractionation, bump de-excitation for G-type stars and serve as an aid for interpreting Ly-$\alpha$ and He$2^3$S observations.}
    \label{fig:distance_variation_all}
\end{figure*}

Employing the spectrum by \cite{bourrier2020} to model HD189733b, we find a nearly perfectly overlapping solution for both thermodynamic models, see Fig. \ref{fig:helium_transits_data_realcomparisons} (second row). Those can however still be distinguished by the scaling factors necessary to fit the line depth - we require scaling the adiabatic model by $\times 1.34$ and the isothermal one by $\times 14$. The former scaling indicates that the model is close to being self-consistent, the latter scaling indicates that either a 10 times higher mass-loss rate is required, or super-solar Helium abundances are present.

The line-broadening however does not quite match. We can gain insight into what might be driving this excessive broadening by comparing the outflow simulation for HD189733b, driven by the \citep{lampon2021} spectrum. We reiterate the results from Table \ref{tab:stellartypes}, which gave much higher mass-loss rates when employing this spectrum. The outflow is also significantly faster at lower radii, compared to when employing the \cite{bourrier2020} spectrum, giving the triplet line an excessive broadening Fig. \ref{fig:helium_transits_data_realcomparisons} (third row) in both the adiabatic and isothermal case. To reach the transit amplitude shown, we employed Helium scaling factors of $\times 0.08$ for the adiabatic and $\times0.4$ for the isothermal simulation. It becomes clear that the HD189733b transit signal is just on the edge of being sensitive to velocity broadening, enhanced by the presence of the population triplet bump. 
While the exact stellar spectrum is not known, we nonetheless interpret this as a possibility to probe the adiabatic physics with HD189733b. While the signal-to-noise on the data for HD189733b is excellent, we speculate whether other K-star hosted planets with faster outflows might be even better suited to distinguish between the two models. We also reiterate the results seen in Fig. \ref{fig:velocity_fractionation_stars}, in that higher-mass stars make it easier for the flow to reach high velocities via their tidal field.


The isothermal simulation does not feature the population bump and can only explain the broadness of the data under physically inconsistent assumptions, i.e. those of even larger mass-loss rates, not supported by the energy input into this planet, and an empirical free parameter, turbulent broadening \citep{lampon2021}.
As mentioned above, a choice of temperature of 10$^4$~K already maximizes the possible thermal broadening. This is therefore the limit of natural uncertainty effect broadening that the isothermal models can achieve.


Contrasting this, our results suggest a more natural explanation, which is that adiabatic cooling produces a helium triplet population bump. In the case of HD189733b, a slightly harder spectrum than that used here would move the population bump to high velocities to explain the line width in HD189733b, at a stellar helium-to-hydrogen ratio. Therefore, the possible signature of adiabatic cooling in HD189733b's outflow, if confirmed, would demonstrate the hydrodynamic outflow on even larger scales than that already demonstrated by metal line transits \citep{dosSantos2023b}.  

Thus, to be fully certain that the excess line width for HD189733b arises from adiabatic cooling, one would need additional observational constraints to fix the velocity in the line-forming region, \citep[see e.g.][]{linssen2023}. For example, Lyman-$\alpha$ transits can constrain the outflow velocity \citep[e.g.][]{Owen2023,Schreyer2024}. {While the current spectral resolution is sufficient for HD189733b,} 
higher spectral resolution observations {for narrower lines} would constrain the exact broadening profile, potentially distinguishing between the two models. 

{Interestingly, our M2 model comparison shows that M2 stars reverse the trend; however, here the impact of velocity broadening is stronger than that caused by the existence of the velocity bump, after scaling the isothermal model by x25 and scaling the adiabatic model by x$0.2$. This seems in contradiction to our velocity-sensitivity analysis in the previous section, but is explained by the fact that to match the mass-loss rates, we had to alter the base density structure of the isothermal simulation. This eliminates the aspect of a clean comparison when matching mass-loss rates; highlight the difficulty on using isothermal models. 
Finally, we remark that in the literature \citep[e.g.][]{lampon2021} rescaling, as we have performed here, has been used to argue for a consistency between photoevaporative and isothermal models. Based on the results in this section, we caution that such conclusions cannot be universal, and as, for example, the agreement in our B20 model for HD189733b should be a coincidence rather than the norm. }

\subsection{Distance variation for all HD189733b analogues}

{Finally, we show a simple experiment in which we vary the planet-star semimajor-axis distance $d$ for all three stellar types that host HD189733b analogues, to highlight some key physics governing the atmospheric escape of these planets. The results for the escape rates in hydrogen, helium, mass flux ratio, and predicted transit depth can be seen in Fig. \ref{fig:distance_variation_all}.}

{Our simulations shows how strongly the mass loss is shaped by the tidal fields of each host star - resulting in the G2 stars triggering the strongest mass-loss rates, due to their stronger tidal field. Furthermore, in most of the distance range probed, the hydrogen loss rate scales as $d^{-3}$, instead of $d^{-2}$, a regime which is only reached for the lowest mass M2 stars, indicating the outflows are tidally dominated.}

{Hydrogen as the main driver of escape and drag therefore also has an important effect on the overall mass-loss rate of helium. Initially, from $d=0.03$AU to $d=0.05$AU the scaling is similar to $d^{-3}$, indicating weak fractionation, whereas at distances larger than $d=0.05$AU, the mass-loss rate fraction starts differing from its primordial value of $0.27$, indicating the onset of fractionation. We do not consider $\rm H_3^{+}$ cooling or other molecular coolants in this work, which might complicate this picture further at distances larger than 0.1 AU \citep{koskinen2007}.}

{The predicted transit signals scale overall similarly as the mass flux ratios, modified by the well-known influence of the spectral type of the host star. However we note that the overall decrease in transit depth with distance is stronger than just the mass flux ratio, as slower outflows also have more time to depopulate the metastable state.
As the determination of mass-loss rates requires a simultaneous fit in multiple tracers, but also knowledge of the relative ratio between those tracers, we show the ratios of neutral Hydrogen and the Helium triplet state $n_{\rm He2^3S}/n_{\rm  H}(r)$ in the bottom row of Fig. \ref{fig:distance_variation_all} as function of planetocentric radius $r$. It is important to note that only for the K2 host the ratio is an approximate constant, whereas for the remaining host types it is a much stronger function of radius. Particularly for the M2 hosts the relatively large importance of large radii might be cut-off in reality due to stellar wind interactions and act to suppress the expected large He$2^3$S transit depth relative to Ly-$\alpha$ or H-$\alpha$ transit depths. We note that this is not necessarily an explanation for the missing transits around M-dwarves - merely an important caveat when using joint constraints from detections of He and H lines. }



\section{Conclusions}

We have used detailed, fully self-consistent radiation hydrodynamic simulations of atmospheric outflows from hot Jupiters to study the impact of different physical assumptions on the helium triplet signature. In particular, we explored the impact of an isothermal assumption commonly used in the modelling of observations. Additionally,  we explored the impact of sometimes omitted helium-level transitions. 
\begin{itemize}
    \item Non-isothermal temperature profiles, key signatures of hydrodynamic photoevaporative outflows, might be detectable via the helium triplet population bump generated in their radial population profile when the outflow cools adiabatically to $\leq 5000K$, which suppresses collisional de-excitation of the helium triplet set. High spectral resolution observations would be required to perform such a study.
    \item The $2^1$S level does not play a role in repopulating $2^3$S. The most up-to-date spontaneous decay rates indicate it rapidly decays to the ground state.
    \item Ground-level de-excitation needs to be considered and can account for about a $10\%$ difference for $\rm K2$ dwarfs and a factor $2$ difference in populations for M-dwarfs, compared to prior work.
    \item Fractionation of neutral helium sets in for M-dwarfs and might lead to a suppression of the He$2^3$S transit signal for distant planets with weak and neutral outflows. This occurs despite He$2^3$S inheriting the momentum of He$^{+}$, which allows its velocity to split from that of ground-state neutral helium.
    Our results for other stellar types indicate that this suppression should not be unique to M-dwarfs, as for close-in planets, K types fractionate more strongly.
    {We caution that the radially increasing $n_{\rm He2^3S}/n_{\rm  H}$ profile, which is seen unresolved in most observations, might be cut off in reality by solar wind-interactions. This might complicate conclusions about fractionation around M-types.}
    \item The helium triplet population bump must be situated at high velocities to help explain anomalous line widths. However, G-type stars suppress this bump through strong photoionization of the triplet population. On the other hand, M-type hosts have weaker tidal fields, lending fast outflows less probable. K-types, possessing flow velocities in excess of $\rm 1\, km/s$ close to the planet, are the best candidates for detecting signatures of adiabatic cooling.
    \item Our standard simulations can explain the large widths of HD189733b's helium triplet line due to adiabatic cooling rather than excess broadening, if HD189733's spectrum is only slightly harder than inferred by \citep{bourrier2020}. However, multiple transit measurements would be required to reduce the noise sufficiently, along with additional tracers so that other scenarios can be conclusively ruled out.
\end{itemize}

\section*{Acknowledgements}
We thank the anonymous referee for helping us to improve the quality of this manuscript. JEO is supported by a Royal Society University Research Fellowship. This project has received funding from the European Research Council (ERC) under the European Union’s Horizon 2020 research and innovation programme (Grant agreement No. 853022). This
work benefited from the 2023 Exoplanet Summer Program in the Other Worlds Laboratory (OWL) at the University of California, Santa Cruz, a program funded by the Heising-Simons Foundation. M.S. thanks Ruth Murray-Clay for making an additional spectrum of HD189733 available.

\section*{Data Availability}

 The {\sc aiolos} code is publically available at \url{https://github.com/Schulik/aiolos}.



\bibliographystyle{mnras}
\bibliography{bib} 




\appendix

\section{Datafits}
\label{sec:appendix_exdexfits}

\begin{figure*}
	\includegraphics[width=1.85\columnwidth]{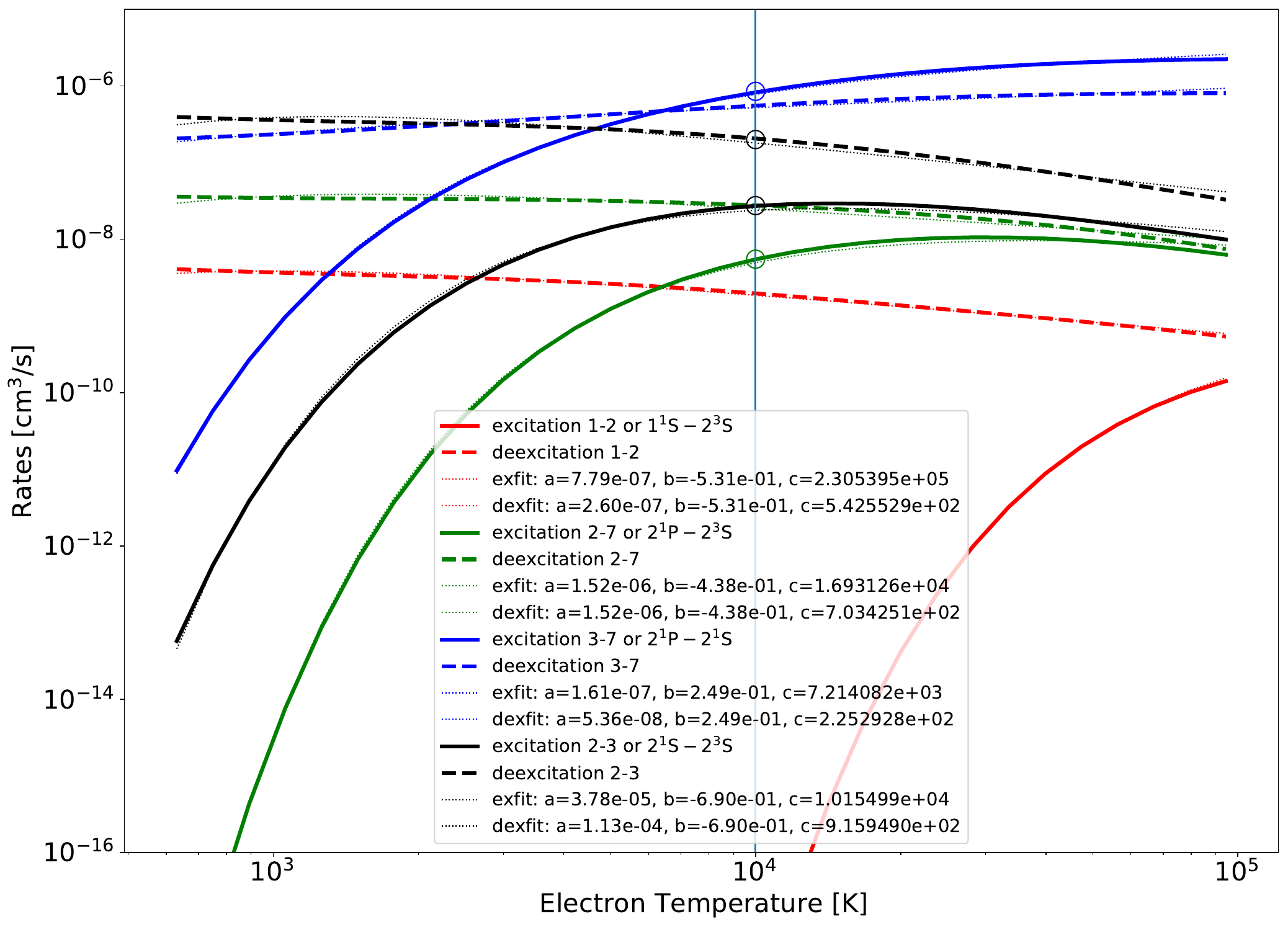}
    \caption{Excitation and deexcitation rates of individual state transitions with thermal electrons at the given electron temperature as extracted from \textsc{Chianti} are shown as thick lines. Our fits over this temperature range are shown as thin lines in comparison, and the corresponding Arrhenius coefficients are shown in the legend. Rates at 10,000~K, which are important for the depopulation of the $2^1$S level, are marked in circles; see text.}
    \label{fig:all_rates}
\end{figure*}

All rates from Table \ref{tab:all_rates} are shown together with our Arrhenius-type fits in Fig. \ref{fig:all_rates}. The hierarchy of rates, which are important to depopulate the $2^1$S level even at $T=10^4$K, are marked with coloured circles.

\section{The new chemistry solver}
\label{sec:appendix_chemsolver}

For this work, we have updated the chemistry solver in \textsc{aiolos}, to be able to handle general photochemical and thermochemical processes. In the following, we describe our approach exemplary on the multispecies ionisation problem.

Following \citep{press2002, tsai2017} The ODE term to be solved for a species $s$ which is being ionised is: 
\begin{align}
    \partial_t n_s = - \Gamma_s\; n_s
    \label{eq:appendix_b1}
\end{align}
Multi-species, multi-band ionisation rates inside of one cell are 
\begin{align}
\Gamma_s &= \sum_b \Gamma_{s, b} \label{eq:appendix_b2} \\
&= \sum_b \frac{\Delta\tau_{sb}}{\Delta\tau_{b, \rm tot}} \Gamma_{b, \rm tot}\\ &= \sum_b \frac{\Delta \tau_{sb}}{\Delta \tau_{b, \rm tot}} 
\frac{F_0 e^{-\tau_{b, \rm tot}}}{\Delta x}
\left(1-e^{-\Delta \tau_{b, \rm tot}} \right) \label{eq:appendix_b4}
\end{align}

The implicit solution for Eqn \ref{eq:appendix_b1} requires knowledge of the analytic derivative (the Jacobian) of Eqn \ref{eq:appendix_b2} w.r.t $n_s$, to advance from timestep $k$ to $k+1$:
\begin{align}
    \frac{n^{k+1}_s - n^{k}_s}{\Delta t} \approx - \left(\Gamma_s \right)^{k} + \left[ \frac{\partial \Gamma_s }{\partial n_s} \right]_{n_s=n^k_s} \left(n^{k+1}_s - n^{k}_s \right) \label{eq:appendix_b5}
\end{align}
which, after some algebra, can be found to be: 
\begin{align}
    \frac{\partial \Gamma_{s,b}}{\partial n_s} = \tilde{F_b} \Delta x \frac{\kappa_{s,b}}{\Delta \tau_b} \left[ \Delta \tau_{s,b} e^{-\Delta \tau_{s,b}} + \left(1-e^{-\Delta \tau_{b}}\right) \left(1- \frac{\Delta \tau_{s,b}}{\Delta \tau_{b}} \right) \right] 
\end{align}

with $\tilde{F} = F_0 / {h\nu \Delta x} \; e^{-\tau}$.
Eqn. \ref{eq:appendix_b4}, in the optically thin limit, takes a more familiar form when considering that $(1-e^{-\Delta \tau_{b,\rm tot}})/\Delta x \approx \kappa_s n_s$, as in e.g. \cite{MurrayClay2009}. Note that our $\Gamma_s$ is not the ionisation rate, but instead, $\Gamma_s/n_s$ takes this place. Solving problem Eqn. \ref{eq:appendix_b2} in the form Eqn. \ref{eq:appendix_b5} is known as semi-implicit approach and features great numerical stability \citep{press2002} via the compensation terms at past time $k$, even when the reaction rates have exponential dependencies.

From the knowledge of those derivatives, the coupling matrix between all species is then constructed for all cells for all photo and thermoreactions, and then solved, giving the new number densities $n^{k+1}_{s}$.  

We note that the above construction is identical for thermochemical reactions, the only difference being that the reaction rate being evaluated as the product $\Gamma_s = \alpha n_s n_{s'} ...$ and hence the partial derivatives generate additional summands for all reactant species per matrix entry.

To compute the momentum and internal energy transfer across the species grid due to species change from $s$ to $s'$, first, the number of density changes is computed. During this step, all changes of number density per reaction r $\frac{\partial n}{\partial t}^r_{s\rightarrow s'}$ are tracked and an implicit matrix system analogous to Eqns. 49-56 in \citep{schulik2023} is solved to transport momentum and internal energy. As example, for the momentum for every simulation cell, the system 
\begin{align}
    m_s\frac{(n v)^{k+1} - (n v)^k}{\Delta t} =  - \sum_r \frac{\partial n}{\partial t}^{r:s\rightarrow s'} m_s v_s^{k+1} + \sum_{r'} \frac{\partial n}{\partial t}^{r':s'\rightarrow s} m_{s'} v_{s'}^{k+1}
\end{align}
results in a matrix equation 
\begin{align}
    (\rho u)^{k+1}_s \left( 1 + \frac{\Delta t}{n^{k+1}_s} \left(\sum_r \frac{\partial n}{\partial t}^{r:s\rightarrow s'} \right) \right) &- \sum_{r'} (\rho u)_s^{k+1} \frac{m_{s'}}{m_s} \frac{1}{n^{k+1}_{s'}}\frac{\partial n}{\partial t}^{r':s'\rightarrow s} \nonumber \\ 
    &= (\rho u)^k_s
\end{align}

which is solved for the momenta $ (\rho_s u_s)^{k+1} = m_s(vn)_s^{k+1}$ for all species, out of which the velocity is calculated, due to $n^{k+1}_s$ being known from the chemistry substep.
The notation $r:s\rightarrow s'$ and the indices $r$ and $r'$ only serve to distinguish between ingoing and outgoing momentum reactions.

This has to be performed in an analogous way for the transport of internal energy between species, which due to their differences in heat capacity, can attain different temperatures after the chemistry step. In practice we however average our temperatures to the mean temperature in this work.

\subsection{Heating and benchmarking of HD209458b}

The heating rates associated with those ionization and dissociation processes need to be computed by re-running a similar loop, while considering the correct band edges for each process for each species. Here, no species change needs to be taken into account, so the previous semi-implicit formalism can be dropped, and we compute the heating rates from the known, updated rates $n^{k+1}_s$ and couple those rates into the pre-existing temperature solver.
Similar to our original, bolometric heating scheme \citep{schulik2023}, we compute a the heating rate in band $b$ in cell $i$,  $\Delta S^b_i$ via
\begin{align}
    \Delta S^b_i = \frac{1}{4} F_i^b \frac{1-\exp{(-\Delta \tau^b_i)}}{\Delta x}
\end{align}
where $F_{i}^b$ is the high-energy radiation reaching cell $i$, and the relevant optical depth inside the cell is the sum over all processes $r$ and opacity carriers $n^r$ contributing to absorption in this band
\begin{align}
    \tau^b_i = \Delta x \sum^r n^r_i \kappa^{b,r}
\end{align}
where $n^r$ is the number density of the dissociated reactant species, and $\kappa^{b,r}$ is the opacity for this process in this band. Running the loop over $b$ initially, avoids double counting of photons. The energy quant $\Delta S^b_i$ represents the energy of all absorbed photons in that cell, but the energy  available per process is proportional to the relative optical depth fraction of the reactant $\tau^r/\tau^r_{b, \rm tot} =  n^r_i \kappa^{b,r} / \sum^r n^r_i \kappa^{b,r}$ and the ratio of photon energy $E^b$ to the threshold energy for the process $E^r_{\rm thr}$, which is $1-E^r_{\rm thr}/E^b$, also known as efficiency factor. The energy subquant

\begin{align}
    \Delta S^{'b,r}_i = \Delta S^b_i \; \frac{\tau^r}{\tau_{\rm tot}} \; \left( 1 - \frac{E^r_{\rm thr}}{E^b} \right)
\end{align}

is then distributed into the product species so that ultimately the heating rate in each species becomes 
\begin{align}
    \Delta S^{'b,r}_{i,s} = \Delta S^{'b,r}_i \; \omega^r_s
\end{align}
where $\omega^r_b$ is an empirical number that is constructed per process, which guarantees that low-mass photodissociation or photoionization products gain most of the energy, and all its constituents add up to $1$. $\omega^r_s$ scales with the species mass $m_s$ as $1/m_s$, and for each process $\sum^s \omega^r_s = 1$. 

This formulation guarantees that we can accommodate multiple different efficiencies of different processes in the same radiation band. It is further a generalization of our approach in \citep{schulik2023} which is able to accommodate all photodissociation and photoionisation phenomena.

We further added a simple masking matrix to avoid numerical mixing of bolometric, non-species changing heating and high-energy species-changing heating when computing the radiation transport.

{The choice of photon energy $E^b$ within a coarse band is somewhat arbitrary. Currently, we employ $E^b=0.99 E^{b-1/2}$, where $E^{b-1/2}$ is the energy at the band interface. This choice guarantees that the heating efficiency for very coarse settings such as $n_b=1$ remains at a finite value. However, setting $E^b$ in the logarithmic band centre can significantly affect the mass loss rates, depending on the coarseness of the wavelength grid. In this work, we used $n_b=22$ in the range $\lambda\in [12,260]$ nm, in which case the mass loss rates can drop by up to 30$\%$ for HD189733b and $\sim8\%$ for HD209458b for this alternative choice of $E^b$.}

{As a benchmark planet, HD209458b can be used to compare our mass-loss rates to those of previous work. \citep{salz2016} find mass loss rates of about $2 \times 10^{10}\,{\rm g~s^{-1}}$ using their own spectrum of HD209458. \cite{yelle2004, garciamunoz2007, koskinen2013} find $5-6 \times 10^{10}\,{\rm g\,s^{-1}}$ using the solar spectrum, and \cite{guo2013} find $5-20 \times 10^{10}\,{\rm g\,s^{-1}}$ for hydrogen using the solar spectrum. As shown in Tab. \ref{tab:stellartypes} we find $\rm 9\times 10^{10}\,g\,s^{-1}$ for hydrogen when using the solar spectrum and around $\rm 7.5 \times 10^{10}\,g\,s^{-1}$ when using the \cite{salz2016} spectrum for HD209458. Our conclusion is that our simulations are on the high end of the theoretical estimates among the various codes, even with different spectra for the same star taken into account \citep{daria2024}. In \citep{xing2023} fractionation in a similar multi-species hydrodynamic framework is discussed. Those authors find a weakly fractionating flow with a mass-loss rate of $\rm 2\times10^{10} g\,s^{-1}$, although they use an efficiency parameter to regulate their flow heating.
Plausible reasons for this are differences in the treatment of multi-band heating compared to \citep{schulik2023}, which showed no such discrepancy, or the fact that we neglect secondary ionizations. \footnote{{The latter process refers to the drop in heating efficiency due to multiple-ionization events of high-energy photoelectrons \citep{garciamunoz2023}, not the doubly ionized $\rm He^{++}$, which is included in our simulations.}}}

\section{Numerical hydrostatic multispecies drift and a simple stabilization approach}\label{sec:appendix_drift}

Our multispecies formulation, in which we solve one Euler equation per species, causes a new kind of unphysical numerical diffusion to occur. 
To illustrate this, we want to qualitatively investigate the intercell fluxes at a production front for two different species, i.e. near the location in which abrupt, larger pressure gradients $dp/dr>0$ are generated.
In order to illustrate the numerical origin of this phenomenon, we consider Roe's approximate linear solution to the full nonlinear problem of the Euler equations \citep{toro2009} in 1-D, typically written as
\begin{align}
    \vec{F}_{i+\frac{1}{2}} = \frac{1}{2} \left( \vec{F}_L + \vec{F}_R  \right) - \frac{1}{2} \sum_p |\lambda_p| \alpha_p \vec{k}_p
    \label{eq:appendix_numericalflux}
\end{align}
where $\vec{F}_{i+\frac{1}{2}}$ is the desired intercell flux between cells $i$ and $i+1$, $F_{L,R}$ are the analytic left and right fluxes computed from the same cells, $ |\lambda_p|$ are the Eigenvalues of the Euler system, $\alpha_p$ the wave strengths and $\vec{k}_p$ are the Eigenvectors of the same system. The flux is a vector of conserved quantities $\rho, \rho u, E$.
We focus on the mass component of this flux, $F_{i+\frac{1}{2}, \rho}$ , and evaluate it close to hydrostatic equilibrium, i.e. $u \ll c$ and $ \rho c \Delta u \ll \Delta p$, where $c=\sqrt{\gamma \frac{p}{\rho}}$ is the sound speed and $\Delta u$, $\Delta p$ are the jumps between the initial left and right data states. The mass flux component then evaluates to 
\begin{align}
    F_{i+\frac{1}{2}, \rho} &\approx \frac{1}{2}\left( (\rho u)_L + (\rho u)_R \right) - \frac{1}{2} \frac{\Delta p}{c} \nonumber \\
    &\approx - \frac{1}{2} \frac{\Delta p}{c}
\end{align}
In other words, the hydrostatic mass-flux gains a pressure-dependent term, not contained in the original, unsolved system. Mathematically, this behaviour stems from the fact that the solved system anticipates compressional effects.
Now consider the situation in which a species' partial pressure $p_s \ll p_{\rm tot}$ and $\Delta p_s \approx p_s$ in a hydrostatic region. 
Then 
\begin{align}
    \Delta p_s/c_s = n_s/ \sqrt{m_s} \sqrt{\frac{1}{2} \frac{\gamma_s-1}{\gamma_s} f R_{\rm gas}}
    \label{eq:appendix_driftterm}
\end{align}
where $n_s$ is the number density, $m_s$ is the particle mass, $f$ are the microscopic degrees of freedom, and $R_{\rm gas}$ is the universal gas constant. From Eqn. \ref{eq:appendix_driftterm} it becomes obvious that species then suffer from nonzero, and species-mass dependent drift, which would not appear in single-fluid multi-species codes \citep[e.g.][]{MurrayClay2009, erkaev2007, salz2016, johnstone2018, Caldiroli2021} or finite-difference based multi-fluid codes \citep{sakata2024}. The fact that we used the Roe solver for this illustrative analysis, but use the HLLC in our simulation practice is irrelevant, as any solution to the Riemann problem must reproduce this drift.

%

More dramatically, this drift phenomenon happens despite strong collisional momentum exchange at high densities, which forces all species to the same velocity.
This affects species produced at ionization fronts, or in local maxima (such as $\rm H_3^{+}$), which will drop throughout the atmosphere and cause numerical instabilities through wave emanation, adiabatic heating spikes and breaching charge neutrality.

\begin{figure}
  \includegraphics[width=1.1\columnwidth]{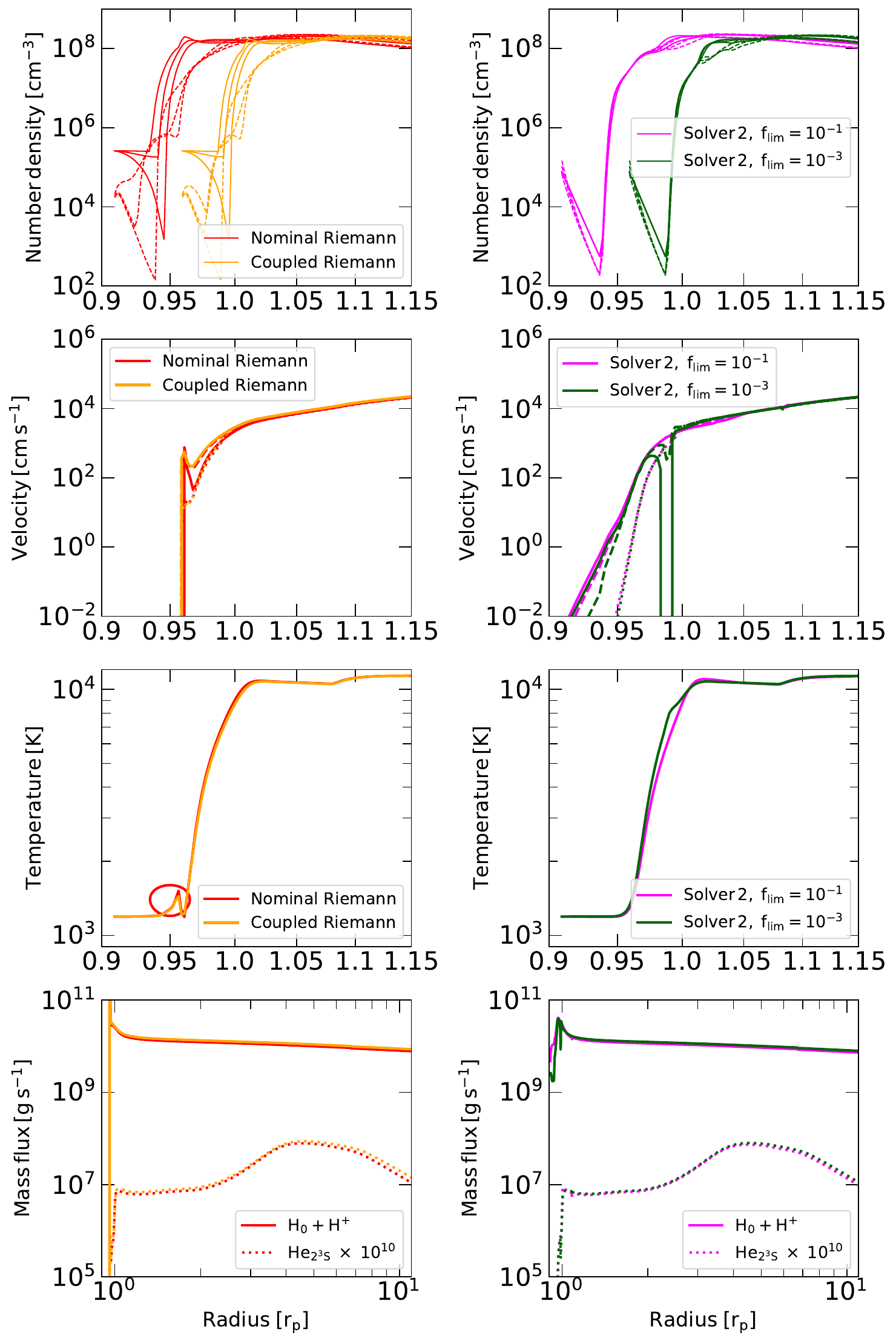}
    \caption{ A comparison of the performances of the original Riemann solver (left column) with our modified solver (right column). The top panels shows electrons (solid lines) and protons (dashed lines) to show the how charge neutrality is (not) kept, and the particle drift. Similar simulations are offset with respect to each other to show the evolution from $1-3$ $\times10^5$s. The panels below show the electron velocities (solid), proton velocities (dashed), $\rm He_{2^3S}$ velocities (dotted), and the averaged particle Temperatures after $3\times10^5$s are below this. The bottom panels show the escape rates in the dominant regime ($f>f_{\rm lim}$) for hydrogen and in the trace regime ($f\ll f_{\rm lim}$) for $\rm He_{2^3 S}$. Those simulations have a very low helium content to show the correctness of the trace regime. Note the different x-axis ranges for mass-loss rates and all other panels. The nominal Riemann solvers show multiple problems: Charge separation occurs and the initial ionization front discontinuity is smeared over time, even in the cases for which we guaranteed all species velocities to be equal (coupled simulations). The tiny temperature bump at the base of the temperature solution for the Riemann solvers (red circle) results from multi-species oscillations and grows over time, destabilizing the solution and shuts off the flow of trace species. All those problems are absent in our modified Riemann solver, when $f_{\rm lim}=10^{-1}$. Lower values of $f_{\rm lim}=10^{-3}$ move the solutions back to being similar to the original Riemann solutions. }
    \label{fig:appendix_riemann_figure}
\end{figure}

To illustrate the origin of this effect further, we enforced all species velocities $v_s$ to attain the same coupled mean $v_{\rm avg}$ after every substep of our algorithm, also testing first and second order Riemann integration. Despite seeding an outflow, electrons and protons started dropping and diffusing, see Fig. \ref{fig:appendix_riemann_figure}(top left).

A simple and general conceptual solution to this problem was to enforce trace species to be solved via a pure advection equation, ignoring its own, now negligible pressure gradients in Eqn. \ref{eq:appendix_numericalflux}, while retaining the pressure terms when they become important for the acceleration of the overall bulk fluid.

We implemented this stabilization scheme by passing the fraction of partial to total pressure $f_s = p_s/p_{\rm tot}$ to the Riemann solver $\Re(\Delta \rho_s, \Delta v_s, \Delta p_s)$ as prefactor in front of the pressure terms, i.e. we solve $\Re(\Delta \rho_s, \Delta v_s, f_{\rm R} \Delta p_s)$, with a multiplicative parameter $f_{\rm R} (f_s)$. This eliminates the numerical drift, but one has to be careful to retain the pressure-related acceleration correctly. Gravity has to be scaled with $f_{\rm R} $ for this to work as well. 
In our numerical experiments we find that in order for the acceleration profiles and hence the mass-loss rates to be identical to the original solver, one needs to ascertain the existence of a region of partial pressure space in which $f_{\rm R}=1$ is passed to the Riemann solver. We therefore introduce a free limiting parameter $f_{\rm lim}$, and pass $f_s=1$ to the Riemann solver if $f_{\rm R} > f_{\rm lim}$, otherwise we pass $f_{s}$ to a smoothing function, so that $f_{\rm R}$ converges to $f_{\rm R} \propto f_s$ for low values thereof and initiates this descent at $f_{s}=f_{\rm lim}$. Specifically we choose

\begin{align}
    f_{\rm R}  = \begin{cases}
1 & f_s \geq f_{\rm lim} \\
 1-\exp{(-f_{s}/f_{\rm lim})} & f_s < f_{\rm lim}
\end{cases} 
\label{eq:numerical_params}
\end{align}

In the limit of $f_s\rightarrow 0$ this approach then successfully reduces all trace particles to purely advected tracers, whereas all species in $f_s \rightarrow 1$ drive the hydrodynamics as they should. This eliminates a number of numerical multi-species problems, including the anomalous drift, from our original formulation and gives close results in terms of mass loss rates, temperature structure and velocity, see Fig. \ref{fig:appendix_riemann_figure}.

From experimenting with the parameter $f_{\rm lim}$, we recommend to use $f_{\rm lim}=10^{-1}$. This means that any species whose partial pressure is 10$\%$ or less of the total pressure, will be gradually transformed into a trace species. All species with higher partial pressure than that contribute to the acceleration of the bulk gas normally. Chosing lower values, such as $f_{\rm lim}=10^{-3}$ lets velocity artifacts reappear, albeit the numerical drift of the ionization front remains absent.

Care needs to be taken in order not to reduce fractionating traces species to non-fractionating, advected trace particles. We therefore set $f_{\rm lim}=10^{-40}$ beyond the radius of $10^{10}$cm, where fractionation would realistically occur for our hot Jupiter simulations in this work. However, we emphasize this radius will be problem and planet dependent. 

We note that we discovered that this effect was already present in our originally published C2Ray solver in \citep{schulik2023}. However, the C2Ray solver retains information about the absolute state of the ionization front at every timestep, i.e. the ion fraction $X(r)$ is known at every $r$ as function of the ionizing flux. Then the number density of ions and electrons was set to $n_{\rm tot} \,X$ and the number density of neutral hydrogen to $n_{\rm tot} (1-X)$. This led to an erasure of the electron drift at every timestep.
However in our new general chemistry solver, presented in Appendix~\ref{sec:appendix_chemsolver}, we only retain differential information about the time evolution, i.e. there is no general analytic solution known with which to re-correct the drifted particles, and number conservation relations cannot be used to reconstruct an arbitrary number of reactants.

The ultimate origin of this problem lies in our choice of solving a multi-Euler system, and in the fact that we cannot well-balance the drift with the frictional momentum coupling terms. A possible algorithmic future solution to this lies in schemes which well-balance pressure gradients and friction, using pressure and advection-split Riemann solvers similar to \cite{thomann2019} or \citep{padioleau2019}.

%

\bsp	
\label{lastpage}
\end{document}